\documentclass[preprint,12pt]{elsarticle}

\usepackage{subcaption}
\usepackage{url}

\usepackage{amssymb}
\usepackage{amsthm}
\usepackage{amsmath}

\usepackage[pdftex,bookmarks, colorlinks, citecolor =blue, breaklinks]{hyperref}
\usepackage{hyperref}

\usepackage{color}

\usepackage{lineno}

\journal{Physica D: Nonlinear Phenomena}

\begin{document}

\begin{frontmatter}



\title{Using scaling-region distributions to select embedding
parameters}


\author[cu-cs]{Varad Deshmukh\fnref{equal}}
\author[cu-cs]{Robert Meikle\fnref{equal}}
\author[cu-cs,sfi]{Elizabeth Bradley}
\author[cu-math]{James D. Meiss}
\author[asu]{Joshua Garland}
\fntext[equal]{These authors contributed equally to this work.}

\address[cu-cs]{{Department of Computer Science, University of Colorado}, 1111 Engineering Drive Boulder CO 80309 USA}

\address[cu-math]{{Department of Applied Mathematics, University of Colorado}, 1111 Engineering Drive Boulder CO 80309 USA}

\address[sf]{Santa Fe Institute, 1399 Hyde Park Rd. Santa Fe 87501 USA}

\address[asu]{{Center on Narrative, Disinformation and Strategic Influence, Arizona State University},
781 S Terrace Road Tempe AZ 85287 USA}

%
%
%

\begin{abstract}
    Reconstructing state-space dynamics from scalar data using time-delay
    embedding requires choosing values for the delay $\tau$ and the
    dimension $m$.  Both parameters are critical to the success of the
    procedure and neither is easy to formally validate.  While embedding
    theorems do offer formal guidance for these choices, in practice one
    has to resort to heuristics, such as the average mutual information
    (AMI) method of Fraser \& Swinney for $\tau$ or the false near
    neighbor (FNN) method of Kennel {\sl et al.} for $m$.  Best practice
    suggests an iterative approach: one of these heuristics is used to
    make a good first guess for the corresponding free parameter and then
    an ``asymptotic invariant'' approach is then used to firm up its value
    by, e.g., computing the correlation dimension or Lyapunov exponent for
    a range of values and looking for convergence.  This process can be
    subjective, as these computations often involve finding, and fitting a
    line to, a {\sl scaling region} in a plot: a process that is generally
    done by eye and is not immune to confirmation bias.
    Moreover, most of these heuristics do not provide confidence
    intervals, making it difficult to say what ``convergence'' is.  Here,
    we propose an approach that automates the first step, removing the
    subjectivity, and formalizes the second, offering a statistical test
    for convergence.  Our approach rests upon a recently developed method
    for automated scaling-region selection that includes confidence
    intervals on the results.  We demonstrate this methodology by
    selecting values for the embedding dimension for several real and
    simulated dynamical systems.  We compare these results to those
    produced by FNN and validate them against known results---e.g., of the
    correlation dimension---where these are available.  We note that this
    method extends to any free parameter in the theory or practice of
    delay reconstruction.
\end{abstract}

\begin{keyword}
Delay-coordinate embedding \sep Nonlinear time series analysis \sep embedding parameters

\end{keyword}

\end{frontmatter}


\section{Overview}\label{sec:overview}

Delay-coordinate embedding, \cite{packard80,takens} the foundation of
nonlinear time-series analysis, involves constructing $m$-dimensional
vectors $\vec{v}(t)$ from a scalar time series $x(t)$, defined by
\[\vec{v}(t) = [ x(t), x(t-\tau), x(t-2\tau), \ldots, x(t-(m-1)\tau)] \]
for a time-delay $\tau$.
If this is done correctly, the reconstructed dynamics will generically
be topologically conjugate to the underlying dynamics that are
sampled by $x(t)$.

There are two free parameters in this procedure: the delay $\tau$ and
the dimension $m$, both of which are critical to obtain a
proper embedding.  The embedding theorems offer guidance for these
choices, but in practice---when one has a finite number of potentially
noisy data points that are measured with finite precision---it is 
typical to resort to heuristics to choose good parameter values.  
Many good strategies have been proposed for these purposes.  One generally
chooses $\tau$ first, working with some statistic that measures
independence of $\tau$-separated points in the time series.  The first
minimum of a plot of the average mutual information versus $\tau$, as
proposed by Fraser \& Swinney \cite{fraser-swinney}, is perhaps the
most common technique.  Subsequently one proceeds to choose
$m$, using e.g., the false near neighbor (FNN) method of Kennel {\sl
  et al.}~\cite{KBA92} In this approach, embeddings of the data for a
sequence of dimensions $m = \ldots, k, k+1, \ldots$ are used to compute
the nearest neighbor to each point at dimension $k$. A change in the neighbor
relationship---if a neighbor in $k$ dimensions is no longer a neighbor
in $k+1$ dimensions---is taken as an indication that the dynamics had
not been properly \textit{unfolded} with $m=k$ and that $m$ should be increased.

This type of heuristic reasoning is difficult to implement as a formal
computational procedure.  For example, the depth of a minimum in discrete plot 
that is required,
the distance that defines a false neighbor, and the maximum
fraction of FNN that signals a proper unfolding can all be subjective.
In the face of these
uncertainties, best practice suggests an iterative method: one of
these heuristics is used to choose a good first guess for the
corresponding parameter. An ``asymptotic invariant'' approach is then
used to firm up the value.  In this procedure, the value of some
dynamical invariant---e.g., correlation dimension or Lyapunov
exponent---is computed over a parameter range to look for
convergence.  This process can also be subjective, however, since
these computations often involve identifying a {\sl scaling region}.
For example in a plot of the correlation sum or distance growth such
scaling regions are generally selected by eye, a
process that is not immune to confirmation bias.  (Of course, if one
simply fits a line to the full results of the calculation without
regard to the plot shape, the resulting value of the
computed dynamical invariant is typically not correct.)
The notion of convergence with
increasing embedding dimension, too, is problematic: is one
significant figure in the correlation dimension enough?  Or does one
need two?  These issues are exacerbated by the fact that when the
embedding dimension is large, the nearest neighbors tend to be far
away, giving incorrect results \cite{Beyer99, Krakovska15}.  Moreover,
large values of $m$ can introduce spurious effects for data sets that
are small or noisy.

In this paper, we address these subjectivities and informalities using
a recently developed method for automated scaling-region selection
\cite{varad-scaling} that offers statistical confidence intervals on
the results.  A sketch of the algorithm is as follows:
\begin{enumerate}
\item On the two-dimensional plot, perform linear fits to segments 
of the data using every possible combination of left and right endpoints.
\item Calculate a weight for each linear fit that is directly
  proportional to the length of the segment and inversely proportional
  to the square of the least-squares fit error.
\item Using the ensemble of fits, generate a histogram of all slopes, 
  taking into account the calculated weights.
\item Generate a probability distribution function (PDF) of slopes
  from the histogram using a kernel density estimator. The mode of
  this PDF is the most likely estimate of the scaling region slope,
  and its full width at half maximum provides confidence
  bounds.\footnote{The algorithm also returns two additional distributions that provide
  information about the boundaries of the scaling region(s).  The
  approach proposed in this paper does not rely on those distributions.}
\end{enumerate}

This technique can be used as the core of an effective methodology,
described in the following section, for automating the asymptotic
invariant procedure.  The algorithm outlined in the steps above not
only removes the subjective identification and extraction of the
scaling regions; it also gives statistical estimates of convergence.
As we describe below, the latter can be computed using an appropriate
metric on the PDFs.  As a proof of concept for these claims, we apply
this methodology to data from a number of real and simulated dynamical
systems to select values for the embedding dimension.  We then compare
the results---both the embedding dimension and the dynamical
invariants---to those produced by other methods.

While we focus here primarily on estimating $m$, it is easy to use this methodology to
estimate good values for $\tau$---or, indeed, for any parameter in a
procedure for calculating dynamical invariants.
One could also use a
straightforward two-parameter extension of our method to estimate
$m$ and $\tau$ simultaneously, as in~\cite{pecoraUnified}.

\section{Automating the asymptotic invariant procedure}

Our goals in this section are to outline a systematic procedure for
selecting good values for the free parameters in the
delay-reconstruction process and to demonstrate the procedure in the
context of the embedding dimension, $m$.
We do this with several synthetic and real data sets, first
estimating the delay, $\tau$, using the method of Fraser \& Swinney
\cite{fraser-swinney}, then embedding the data for a range of $m$ and
computing the correlation sums using TISEAN
\cite{Hegger:1999yq,tisean-website}.  Using the method of Deshmukh
{\sl et al.}~\cite{varad-scaling} on the resulting plots
and the Wasserstein metric \cite{vallender1974} on the resulting
distributions, we establish the dimension for which the
correlation dimension converges.  These results are compared
to the dimension given by the false near neighbor method
\cite{KBA92}.  We also compare the correlation dimension results
to the known values, where they exist.  Finally, we apply these
ideas to computation of Lyapunov exponents. We will note
that algorithms to compute different dynamical
invariants might work best using different embedding dimensions.

\subsection{Data sets}\label{sec:datasets}

We use four data sets in this work.
\begin{itemize}
\item The $x$ coordinate of a 90,000-point trajectory from the
  canonical Lorenz system \cite{lorenz63}:
	\begin{align}
		\dot{x} &= 10 (y - x), \nonumber \\
		\dot{y} &= x (28 - z) - y, \nonumber \\
		\dot{z} &= xy - \frac{8}{3} z, \nonumber
	\end{align}
with the initial condition $(0, 1, 1.05)$. This is obtained using a
fourth-order Runge-Kutta algorithm for $10^5$ points with the
time step $\Delta t = 0.01$.  We discard the first $10^4$ points to
remove transient behavior and focus on the attractor dynamics.  For this
well-studied system the correlation dimension and
largest Lyapunov exponent are well-known
\cite{sprott2003chaos,wolf1985determining}.

\item The first coordinate of a 990,000-point trajectory from the
14-dimensional Lorenz-96 system \cite{lorenz96Model}:
 \begin{align}
	\frac{dx_k}{dt} = (x_{k+1} - x_{k-2})x_{k-1} - x_k + F 
\end{align}
for $k = 1,\ldots 14$ with $x_{k\pm14} = x_{k}$.
The trajectory for the initial condition $[6, 5, 5, \ldots, 5]$ is obtained 
using the fourth-order Runge-Kutta algorithm with time step $\Delta t = \frac{1}{64}$.
We discard the first $10^4$ points from the million point trajectory to remove the transient.
This example is included because its dynamics are high dimensional.
\item Two 80,000-point data sets from experiments on a Photonic
  Integrated Chip (PIC) distributed feedback laser that was developed
  as part of the European Commission PICASSO project, sampled at 40
  GHz \cite{McMahon}.  These examples are included to validate
  our method on experimental data for which there are established
  values for delay-reconstruction parameters and correlation
  dimension.
\end{itemize}

\subsection{Results}\label{sec:results}

\subsubsection{Correlation Dimension}\label{sec:dcorr}

In this section, we demonstrate how to choose good values of the
embedding dimension, $m$, for the four data sets described in
Section~\ref{sec:datasets} using automated asymptotic invariant
analyses on correlation-sum plots.

\begin{figure*}
  \begin{center}
    		\subfloat[Lorenz-63 time series]{
                  \includegraphics[height=0.25\linewidth]{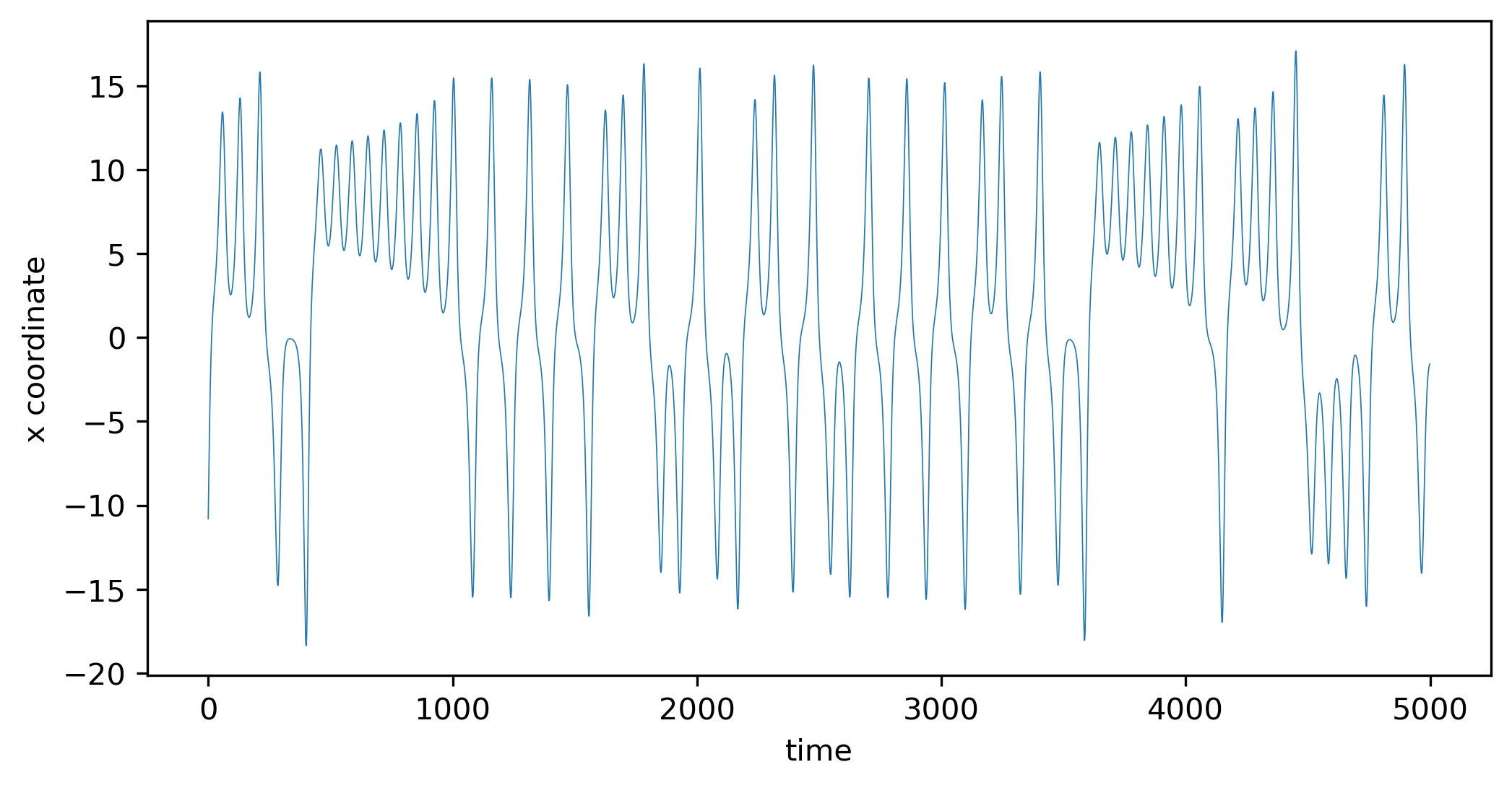}} \\
			\subfloat[Choosing $\tau$: Average mutual information]{
                  \includegraphics[height=0.3\linewidth]{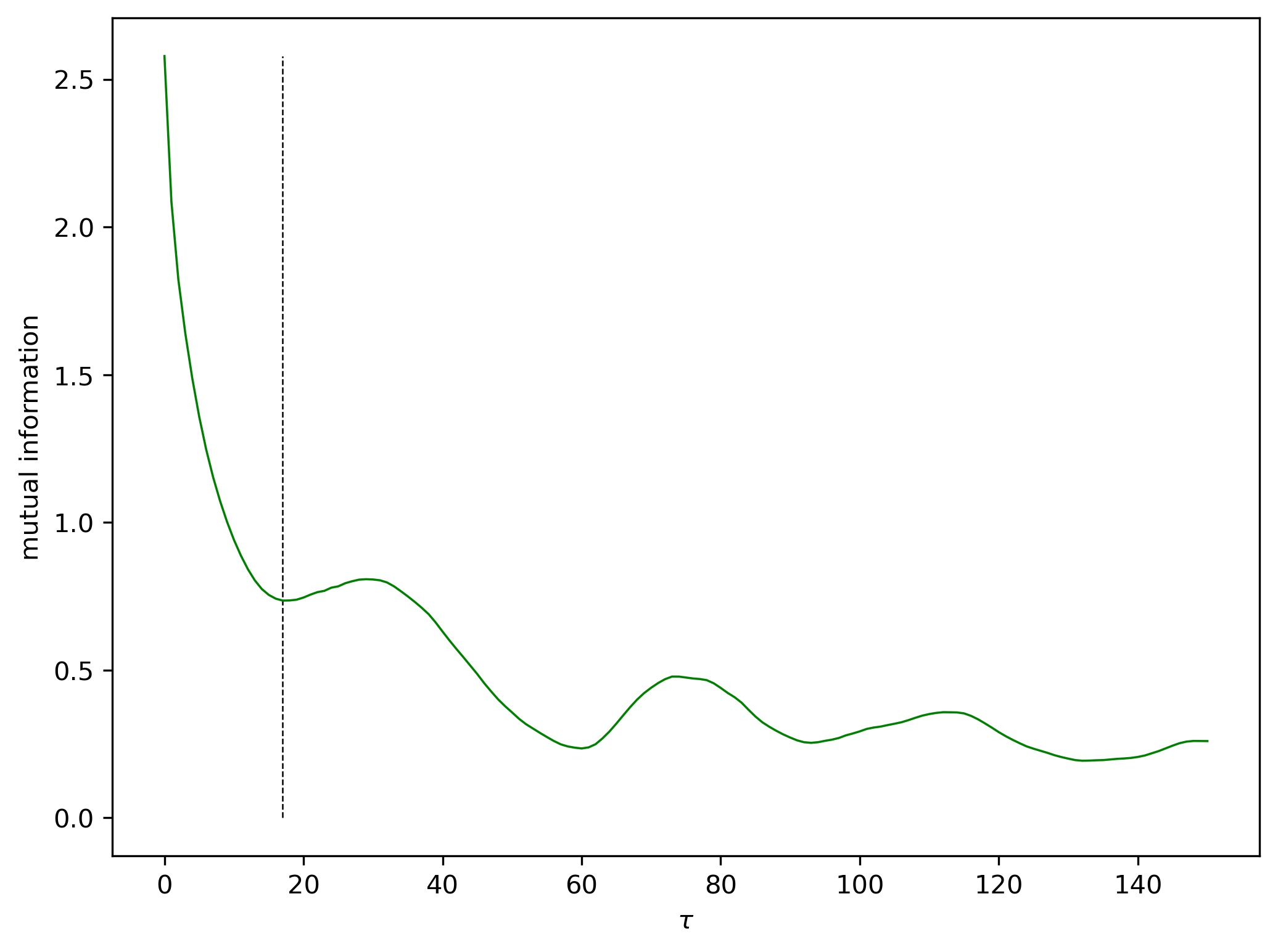}}               
            \subfloat[Choosing $m$: False near neighbor]{
                  \includegraphics[height=0.3\linewidth]{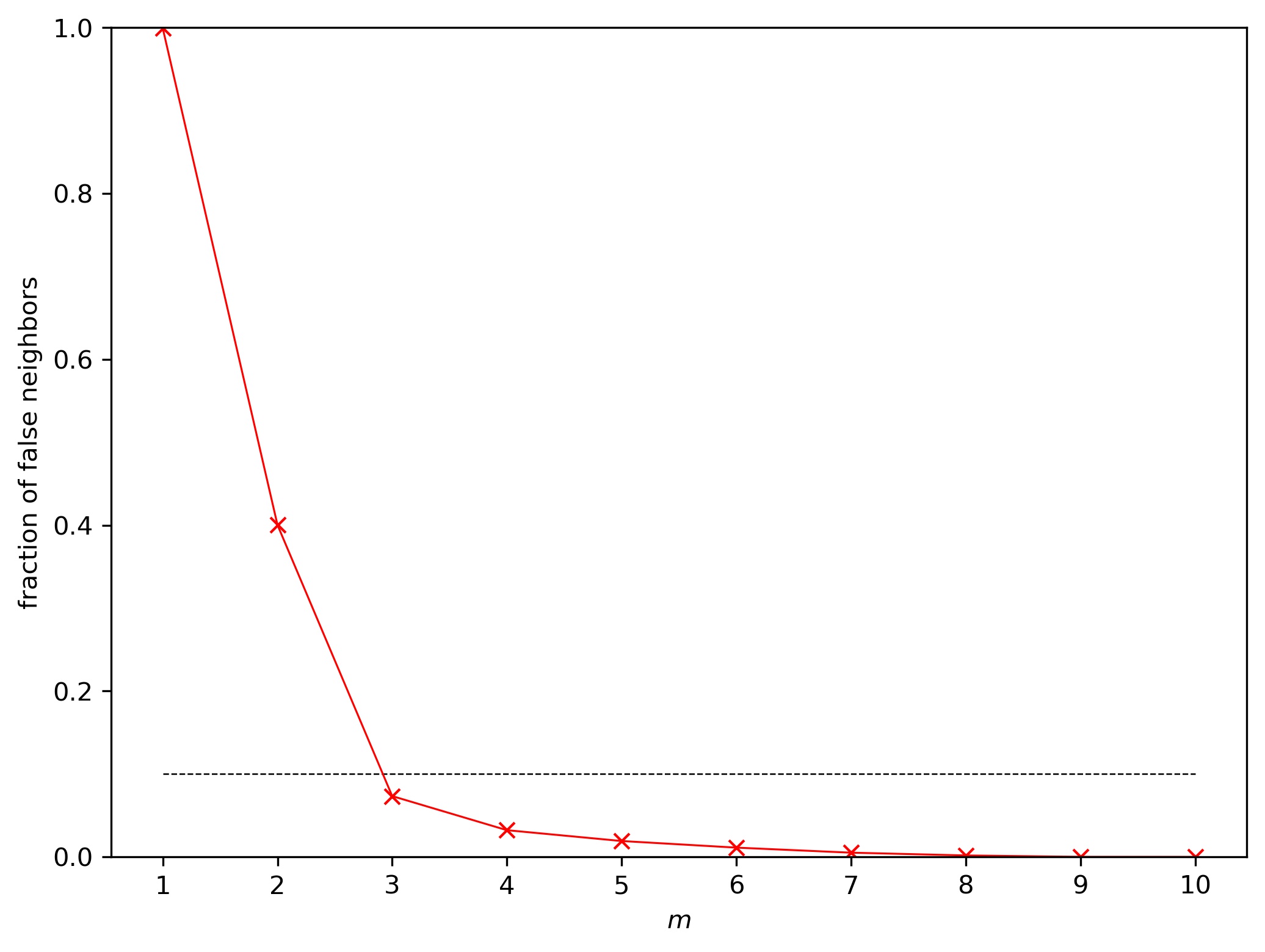}}\\
            \subfloat[Correlation sums]{
                  \includegraphics[height=0.3\linewidth]{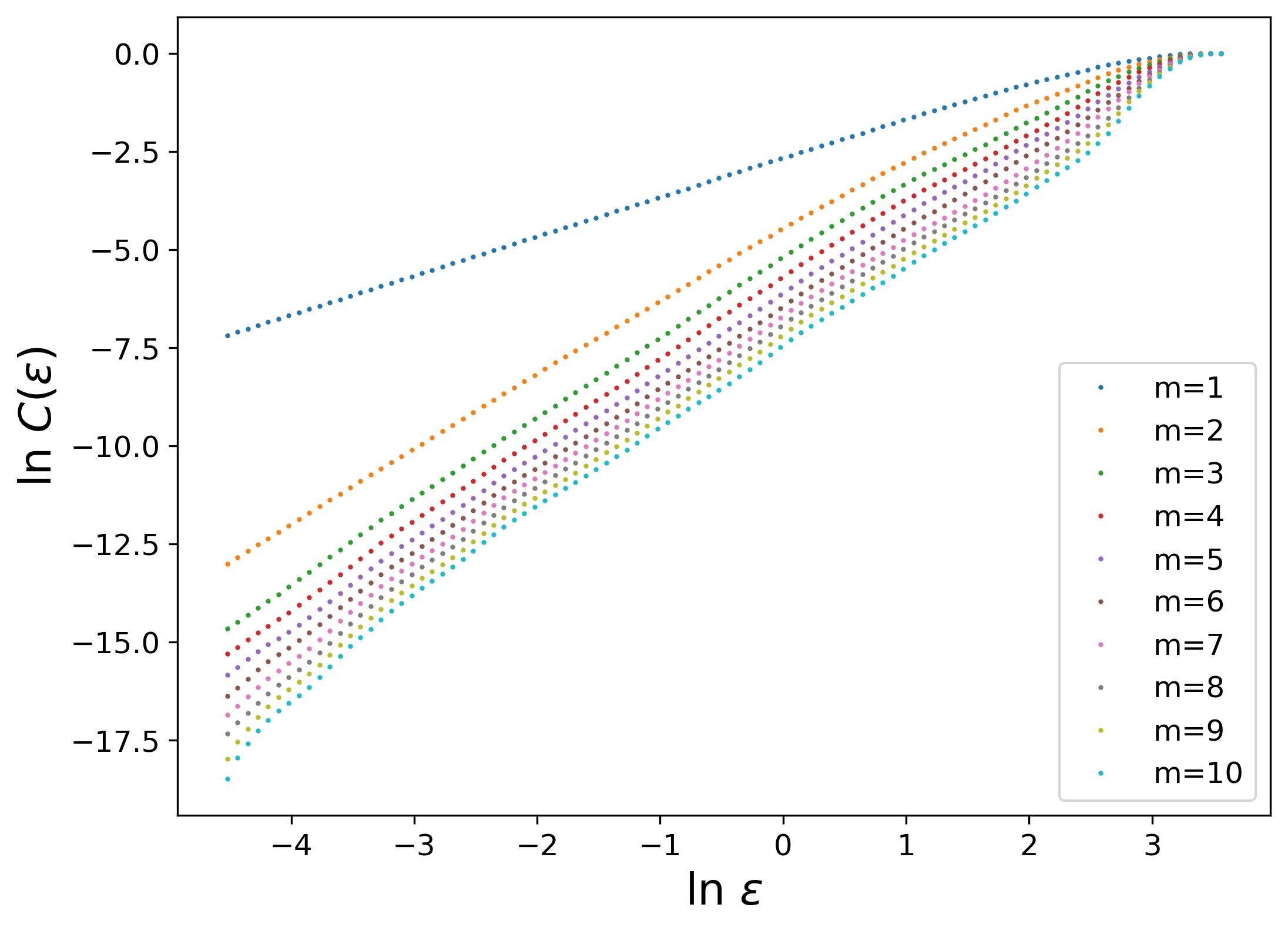}}
            \subfloat[Weighted slope distributions]{
                  \includegraphics[height=0.3\linewidth]{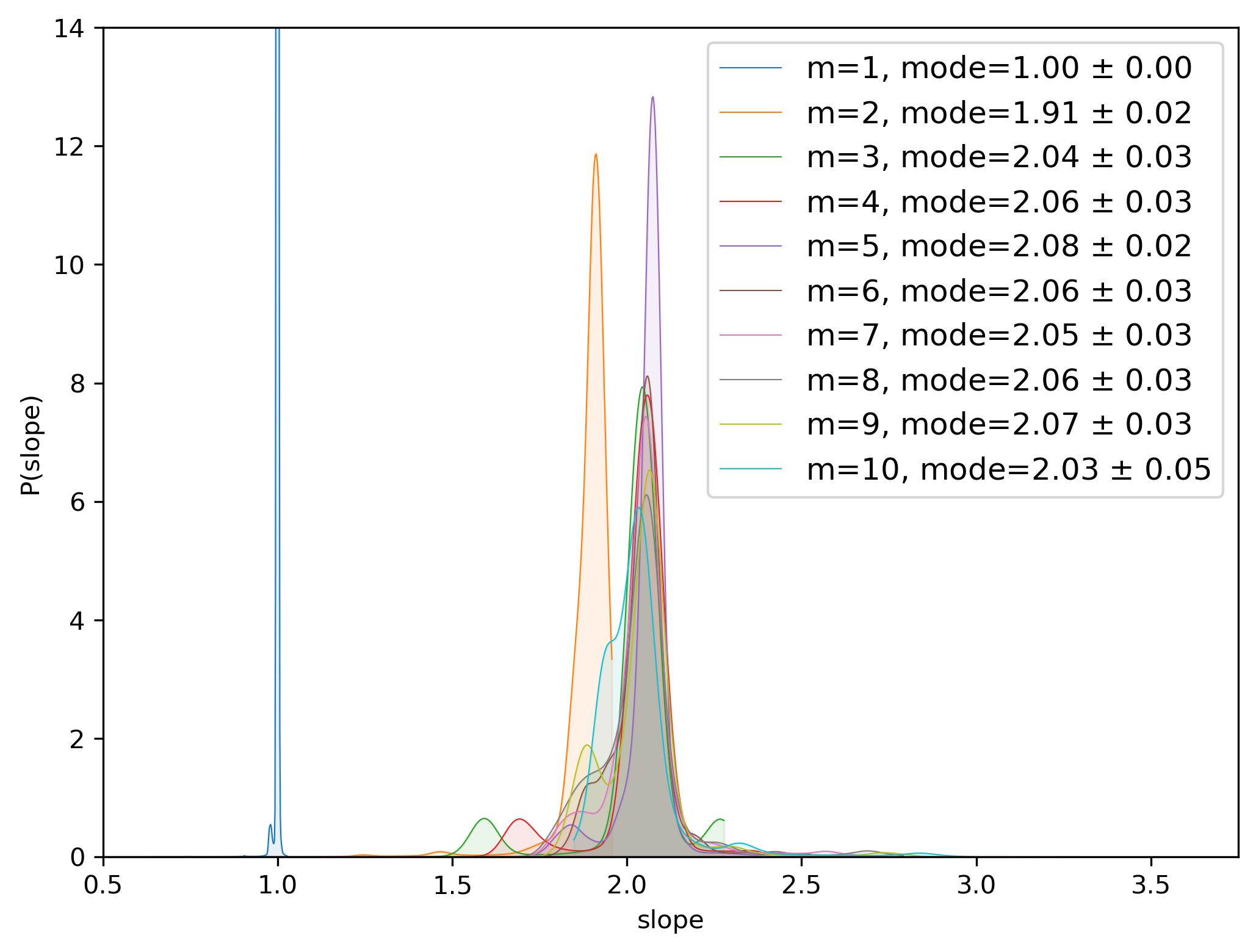}}\\
			\subfloat[Wasserstein distance]{
                     \includegraphics[height=0.3\linewidth]{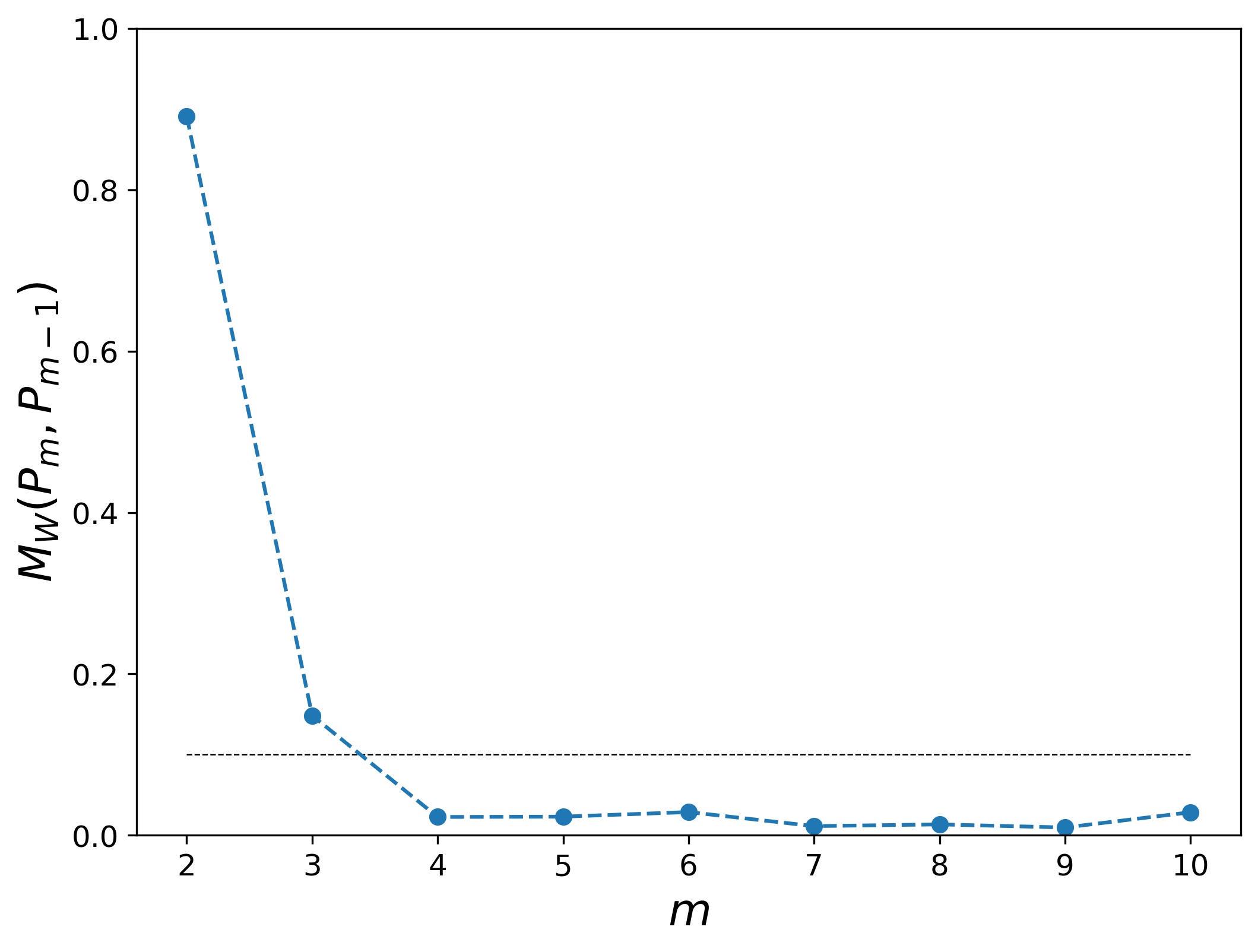}}
            \end{center}
  \caption{Extracting a scaling region using an ensemble-based
    approach (a) Example time series: $x$ coordinate of the Lorenz-63
    system.  (b) Average mutual information as a
    function of $\tau$, computed using TISEAN's {\tt mutual} command.
    (c) Percentage of false near neighbors as a function of the
    embedding dimension $m$, computed with
    $\tau=18$ using TISEAN's {\tt false\_nearest} command; (d)
    Correlation sum plots for embeddings with
    $\tau=18$ and $m=[1,10]$, computed using TISEAN's {\tt d2} command;
    (e) Weighted slope distributions generated from an ensemble of
    fits in different intervals from panel (d).
  (f) Wasserstein distance between successive slope distributions.}
\label{fig:lorenz63-example}
\end{figure*}

The classic Lorenz-63 system is shown in Figure~\ref{fig:lorenz63-example}. 
The first three panels show the standard steps in the
delay-reconstruction process. From the time series, shown in panel
(a), TISEAN's {\tt mutual} command gives the average mutual
information versus $\tau$, shown in panel (b). We select the first minimum
at $\tau=18$ for the rest of the analysis. To estimate the embedding
dimension $m$, we then run TISEAN's {\tt false\_nearest} command;
panel (c) shows the percentage of false near neighbors plotted versus $m$.
Using a 10\% threshold, as is common in practice, the FNN results
suggest $m=3$.\footnote{In this paper we leave
TISEAN's many algorithmic parameters at their default values unless
otherwise mentioned. For Lorenz-63, we increased
the default range of $\tau$ in {\tt mutual} to see the first minimum
in Figure~\ref{fig:lorenz63-example}(b).}

The bottom three panels of Figure~\ref{fig:lorenz63-example}
demonstrate our methodology using the correlation dimension as an
asymptotic invariant.  The correlation sums, $C(\epsilon)$, are found
from TISEAN's {\tt d2} command for a range of $m$ values. Here
$\epsilon$ is the size of the balls used to cover the set during the
calculation of the Grassberger-Procaccia
algorithm \cite{GrassbergerPhysicaD}.
Panel (d) shows $\ln C(\epsilon)$ versus $\ln \epsilon$.
If this plot has a scaling region, its slope is the correlation dimension.

It is common practice to choose the endpoints of a scaling region 
by eye, and then compute the slope using a linear fit.  
In this case, if the slopes were to converge as
$m$ increases, it is thought that the $m$-embedded attractor is properly
unfolded and that the value of the correlation dimension is correct.
Figure~\ref{fig:lorenz63-example}(d) shows clear scaling regions whose
slopes behave as expected: when $m$ is too low, the attractor is not
properly unfolded, so the correlation dimension---i.e., the slopes of
the blue ($m=1$) and orange ($m=2$) traces---is artificially low.  As
$m$ increases, the slopes increase and then appear to converge.

We formalize this procedure using the method of Deshmukh 
{\sl et al.}~\cite{varad-scaling}, which uses slope distributions to identify
scaling regions and the Wasserstein distance to establish convergence
of the distributions with increasing $m$.  As a first step, we compute
potential scaling regions corresponding to an ensemble of intervals
with various values of $\ln(\epsilon)$ for left and right endpoints.
We set the minimum number of points for the fitting interval to be
$10$, but allow all possible combinations otherwise.  This choice is
discussed in \cite{varad-scaling}. For example, in panel (d) there are
$100$ possible selections of endpoints, so using a minimal width of
$10$ gives $4005$ potential scaling regions.  For each $m$ in
Figure~\ref{fig:lorenz63-example}(d), we then generate a distribution
of slopes, $P_m$, from least-squares fits for each interval.  The
goodness of the fit is included by weighting each result by the length
of the fitting interval and inversely by square of the fit error.  We
show kernel density estimates for these distributions in panel (e),
calculated using python's {\tt scipy.stats.gaussian\_kde} command.

The geometry of these distributions brings out the salient information
quite effectively, including both the existence of one or more scaling
regions and their slopes.  Unimodal slope distributions,
as in Figure~\ref{fig:lorenz63-example}(e), suggest the
presence of a single, wide scaling region for {\tt d2}.\footnote
{Note that all distribution plots in this paper have the same vertical
  scale for the purposes of comparison, and may be truncated.}
The mode of $P_m$ is an estimate of the slope of the scaling region
and the width of the distribution around that mode width gives an
indication of precision.  More formally, we calculate a confidence
interval by computing the standard deviation, $\sigma$, of the
ensemble members within the full width at half maximum (FWHM) of the
mode.  For the $m=2$ case (orange), $\sigma =
0.02$, giving the estimated slope $1.92 \pm 0.02$.

If there were \textbf{no} scaling region in the plot, the distribution would be wide
and the corresponding confidence interval large.  For
Figure~\ref{fig:lorenz63-example}, the trajectory samples the
attractor cleanly and thoroughly, resulting in small error estimates.
However, this is not the case for all of the examples
below.  Moreover, if the plot contains {\sl multiple} scaling regions,
the distributions will be multi-modal. This may occur, for example, 
for {\tt d2} when $\epsilon$ is larger than the diameter of the
attractor, or for noisy data when $\epsilon$ is small
\cite{varad-scaling}. The possibility of such multi-modal distributions
is why we use the mode rather than the mean.

The choice of the smallest embedding dimension that gives an accurate
and valid calculation of the correlation dimension is the critical
matter at issue here.  We assert that this $m$ corresponds to the
smallest value for which the slope distributions ``converge.''  In
Figure~\ref{fig:lorenz63-example}(d), this convergence is apparent to
the eye: the $P_1$ (blue) and $P_2$ (orange) distributions
reflect the low correlation dimensions of an incompletely unfolded
attractor; however, the $P_m$ for $m \ge 3$ largely overlap.  This
suggests that $m=3$ or $4$ would be a good choice.

To formalize the notion of convergence, we use the Wasserstein metric
\cite{vallender1974}, $M_{W}$, to compare sequential pairs $P_m$ and
$P_{m-1}$.  As a metric, $M_{W} = 0$ if and only if the distributions
are identical, or---as we are using it for samples---if and only if
the weighted sample values are the same.
Figure~\ref{fig:lorenz63-example}(f) shows $M_W(P_m,P_{m-1})$ for the
Lorenz-63 {\tt d2} slope distributions, calculated using the {\tt
python scipy.stats.wasserstein\_distance} package.  For this
noise-free, low-dimensional case, the distance
$M_W(P_m,P_{m-1})$ montonically decreases with $m$.

For real-valued data, it is known that the $L_1$ Wasserstein distance
for a sample of size $N$ from a distribution approaches zero as
${N^{-1/2}}$ under some technical assumptions \cite{DelBarrio99}.  In
our experiments, $N = \mathcal{O}(10^3)$ is the number of selected left and right endpoint
pairs for the linear fits.  The
theoretical error is also proportional to the width of the PDF, which
in our applications tends to be $\mathcal{O}(1)$. We make the null
hypothesis that the PDFs are the same if
\[
	M_W(P_m, P_{m-1}) \lesssim 0.1 .
\]
In Figure~\ref{fig:lorenz63-example}(f) this threshold, shown as the
dashed line, first occurs at $m=4$
where $M_W(P_4,P_3) = 0.025 < 0.1$, so we choose this embedding
dimension. This then gives $d_2=2.06 \pm 0.03$.

Our approach bears some similarities to other methods for choosing
$m$.  Our procedure employs a threshold that superficially resembles
that used in the FNN method, but the $M_W$ threshold is mathematically
justifiable.  By contrast, there appears to be no such justification
for the selection of a threshold for the percentage of false nearest
neighbors.  The suggestion of~\cite{KBA92} is that ``a physicist
might well choose to accept this threshold to make more efficient
any further computations performed on the data,'' provides a reason
based only on convenience.  Moreover, the
percentages of FNN can vary widely with $\tau$  and
$m$, and also are sensitive to noise~\cite{Krakovska15}.  This further
complicates the selection of a  threshold for the FNN heuristic. 
Similarly, Cao \cite{Cao97Embed} proposes a method to automate the
asymptotic invariant approach by comparing quantities calculated
from embeddings at successive dimensions.  The quantities are derived
from distances between points that are neighbors in space
($E1(d)$) or in time ($E2(d)$). However, the paper does formalize
a threshold on $E1$ and $E2$ to indicate that the correct
embedding dimension has been reached.

\begin{figure*}
 \begin{center}
		\subfloat[Correlation sums]{
      			\includegraphics[height=0.26\linewidth]{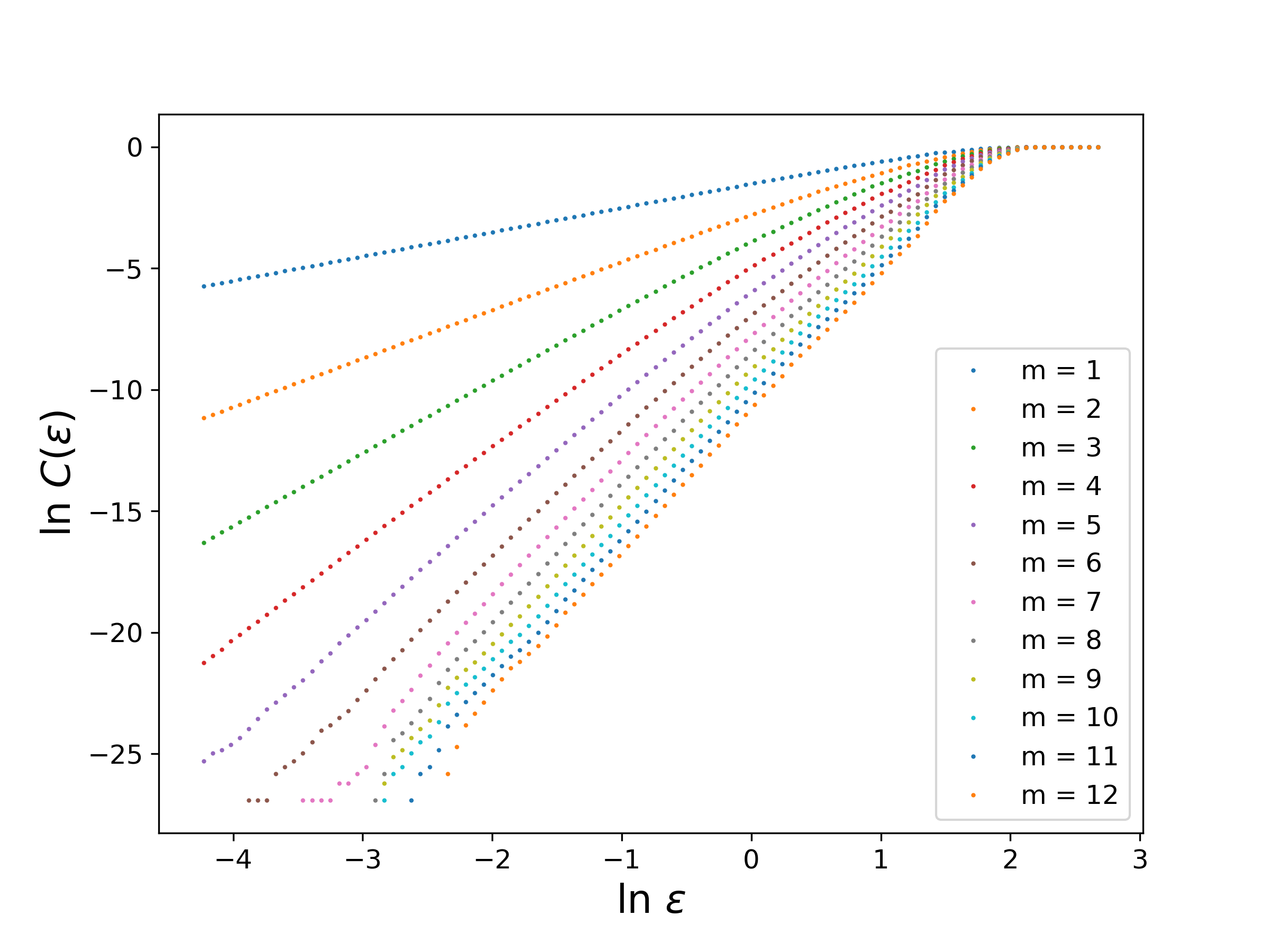} } 
      	\subfloat[Weighted slope distributions]{
                  \includegraphics[height=0.23\linewidth]{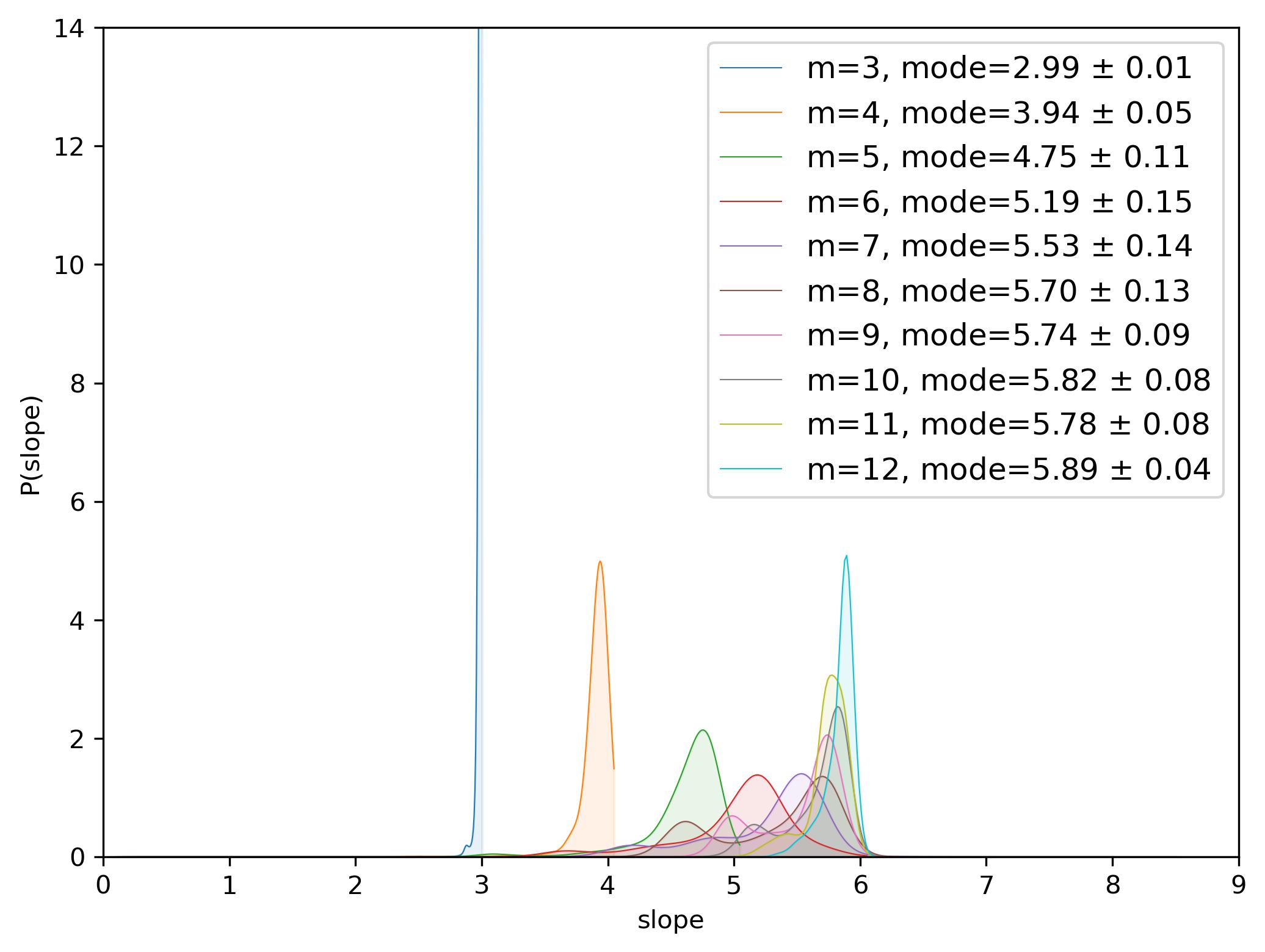}}
		\subfloat[Wasserstein distance]{
                     \includegraphics[height=0.26\linewidth]{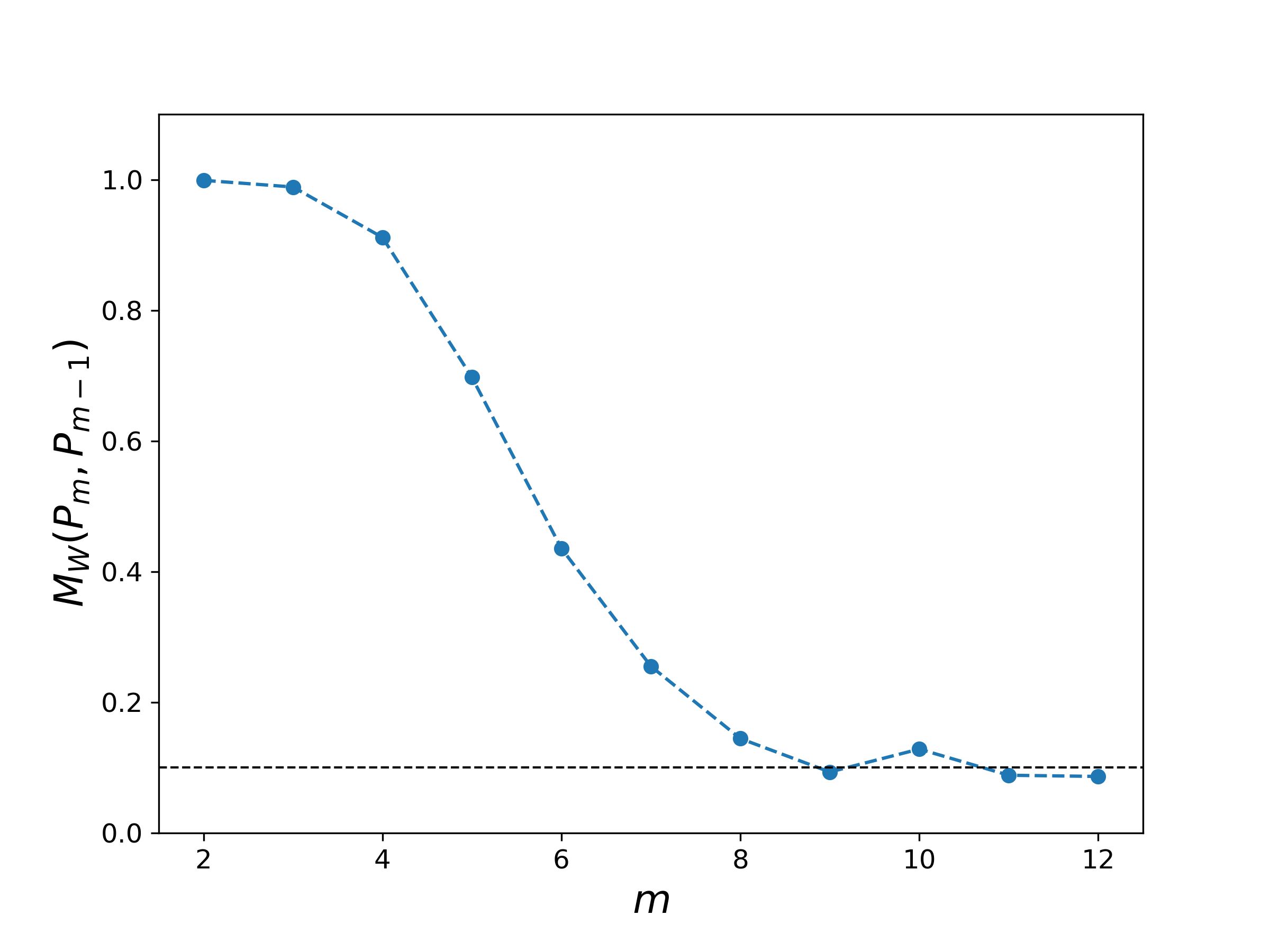}}
\end{center}
  \caption{Lorenz 96 example: (a) Correlation sum plots for embeddings
    for $\tau=23$ and $m=[3,12]$, computed using TISEAN's {\tt d2}
    command. (b) Weighted slope distributions generated from an
    ensemble of fits in different intervals of the plot in (a). (c)
    Convergence of slope distributions.}
\label{fig:lorenz96-example}
\end{figure*}

For the second example, we use the Lorenz-96 trajectory described in
Section~\ref{sec:datasets} to give a time series sampled from an attractor
in a 14D state space.  In this case, AMI (not shown here) does not give a
good estimate for $\tau$ because it has broad, almost-flat region with
a first minimum at $\tau=145$, a value that produces an over-folded
embedding.  Instead, we use the curvature-based heuristic of
\cite{varad-curvature} to select $\tau=23$.
The resulting correlation sums from {\tt TISEAN} for a range of
embedding dimensions are shown in
Figure~\ref{fig:lorenz96-example}(a).
The corresponding slope distributions, panel (b), exhibit the same
behavior as the Lorenz-63 example: they peak at artificially low
slopes when the dimension is too small, and appear to converge with
increasing $m$.  The Wasserstein plot, panel~(c), confirms this and
suggests $m=9$ is sufficient.  This gives $d_2 = 5.74 \pm 0.09$.  

For this data set, the FNN method would require a larger value,
$m=11$, which gives a slightly larger estimate of the
correlation dimension.
The difference between the two estimates stems from what each method
is trying to do.  FNN performs an aggregate calculation of neighbor
relationships across the attractor, with the goal of identifying false
trajectory crossings created by inadequate unfolding.  Elimination of
such crossings is sufficient for computing the correct dimension, but
not necessary \cite{Krakovska15}. By contrast, our method
uses the convergence of the desired invariant as the primary
criterion, which is more appropriate given that this is the goal.

\begin{figure*}
  \begin{center}
    		\subfloat[Laser data set \#1]{
                  \includegraphics[width=0.4\linewidth]{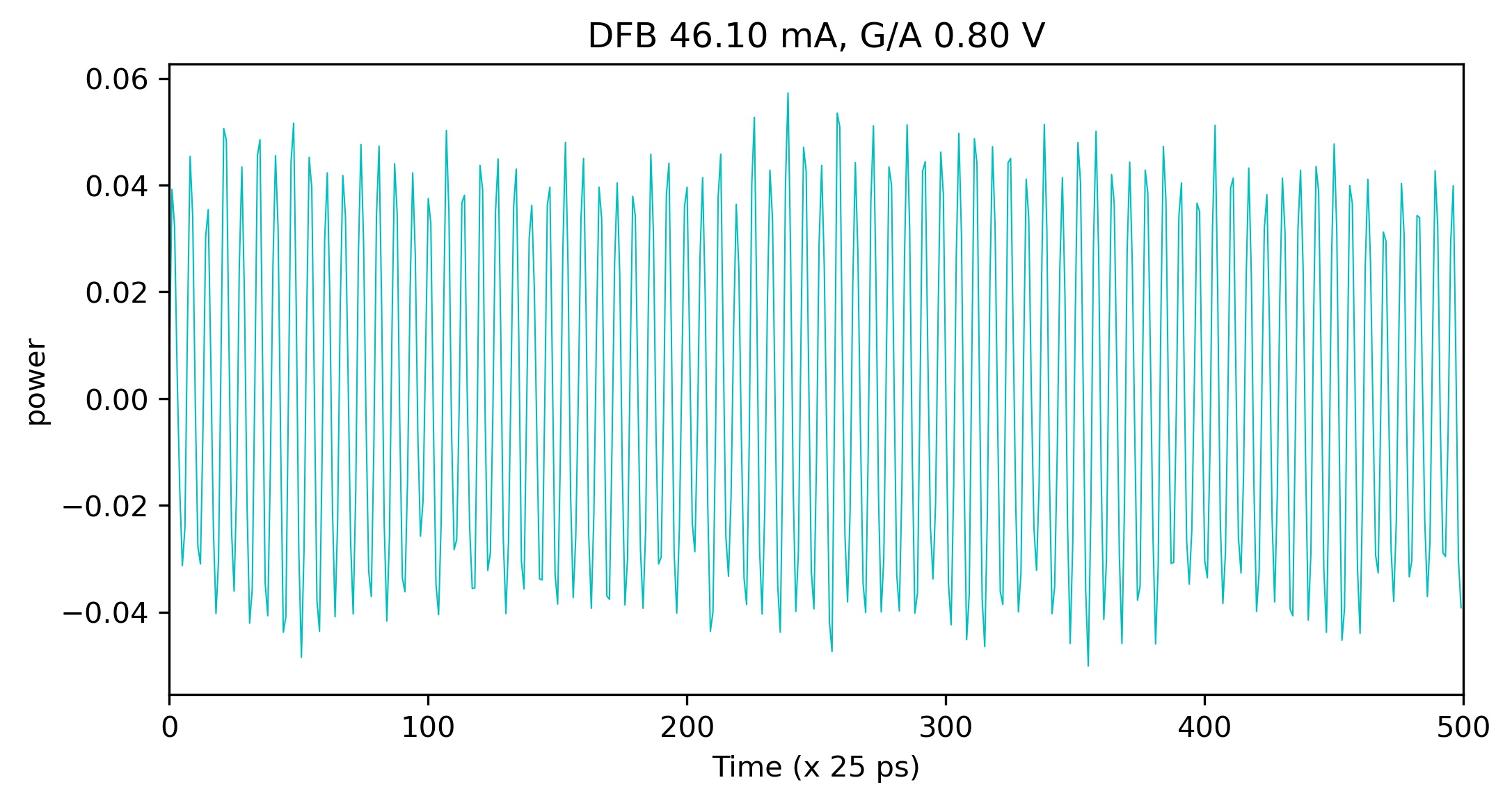}
                } 
            \subfloat[Laser data set \#2]{
                  \includegraphics[width=0.4\linewidth]{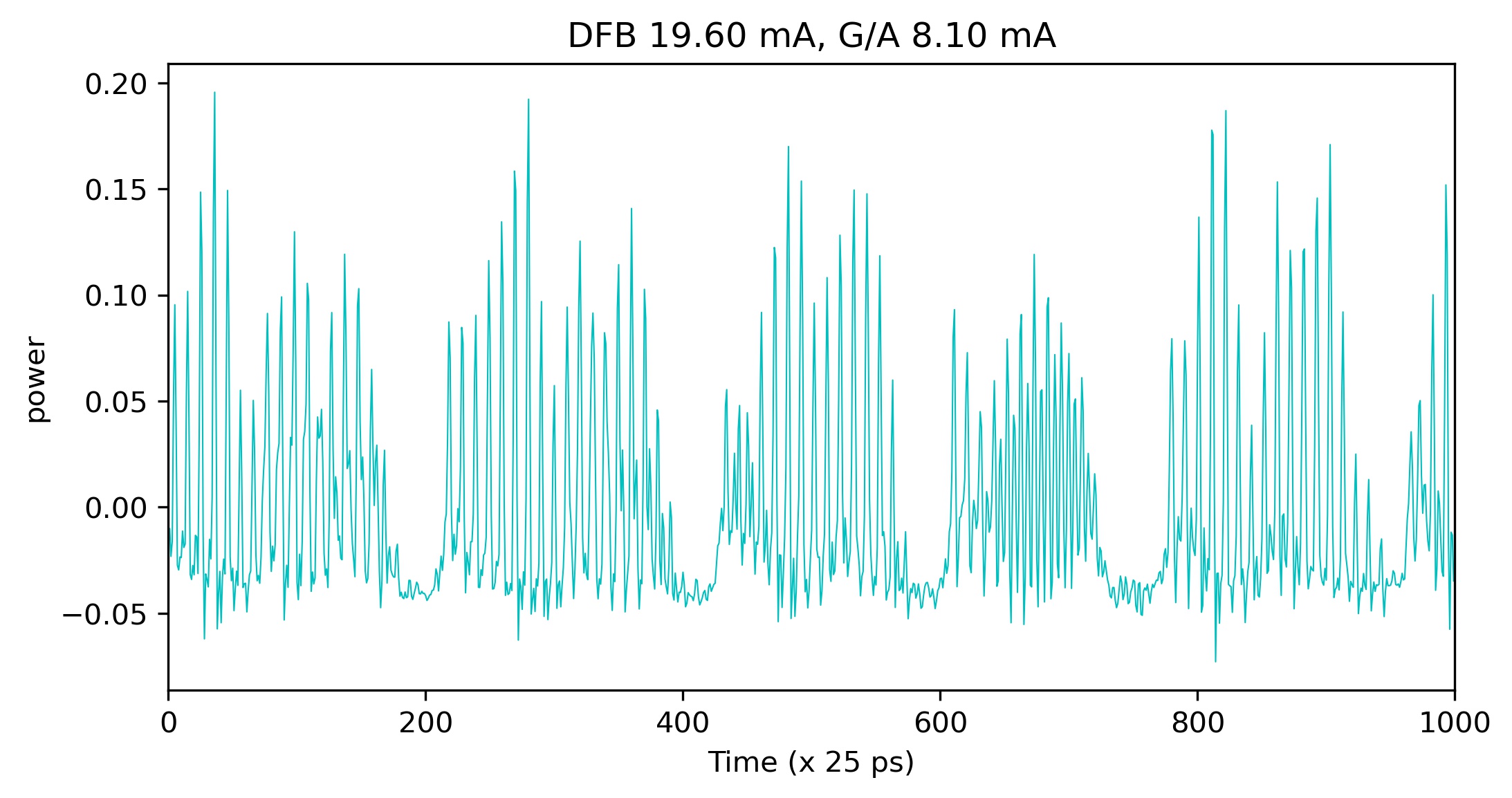}
                } \\ 
            \subfloat[Correlation sums for data set \#1]{
                  \includegraphics[width=0.4\linewidth]{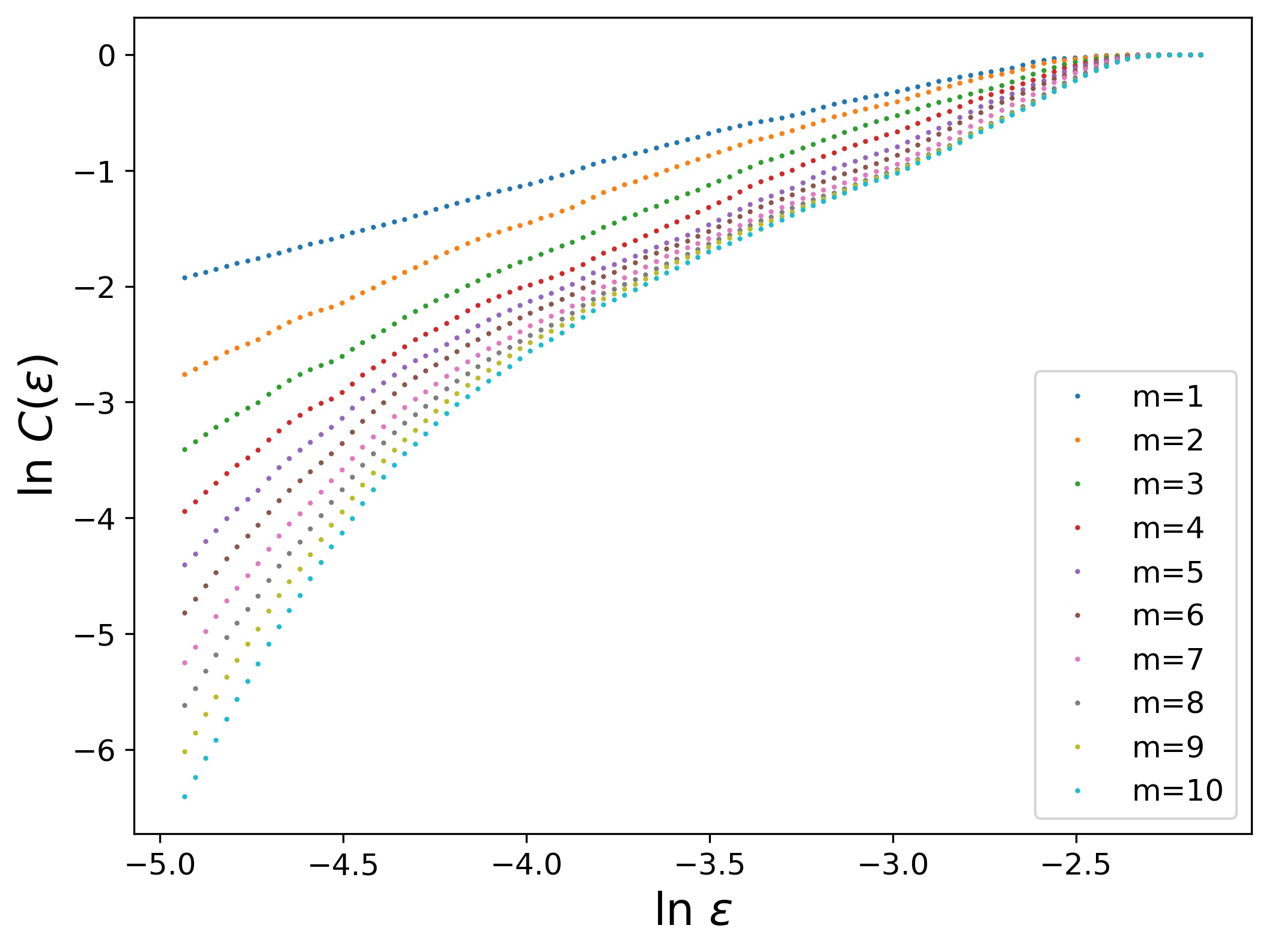}
                } 
            \subfloat[Correlation sums for data set \#2]{
                  \includegraphics[width=0.4\linewidth]{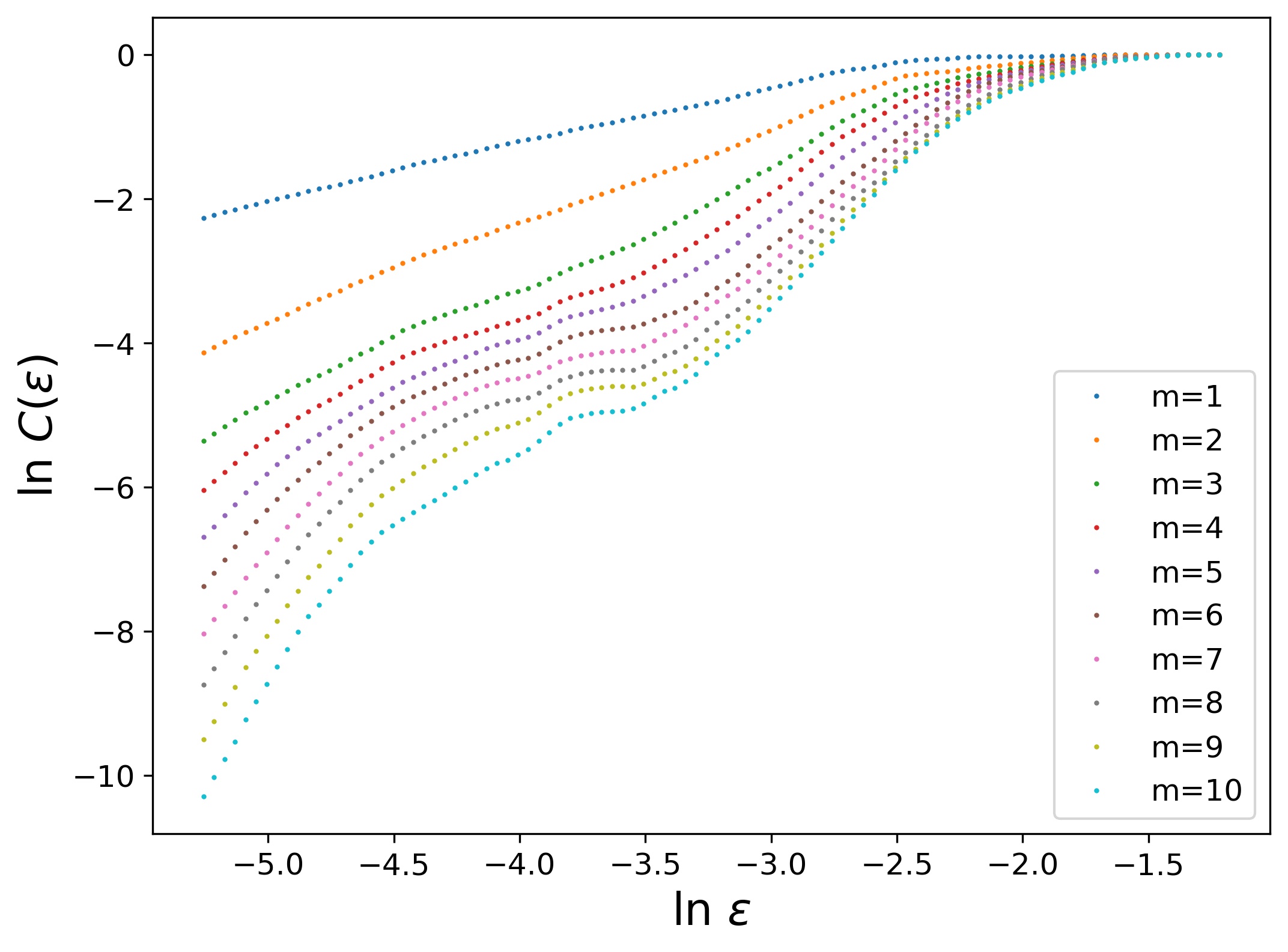}
                } \\
             \subfloat[Slope distributions for data
                  set \#1 ]{
                  \includegraphics[width=0.4\linewidth]{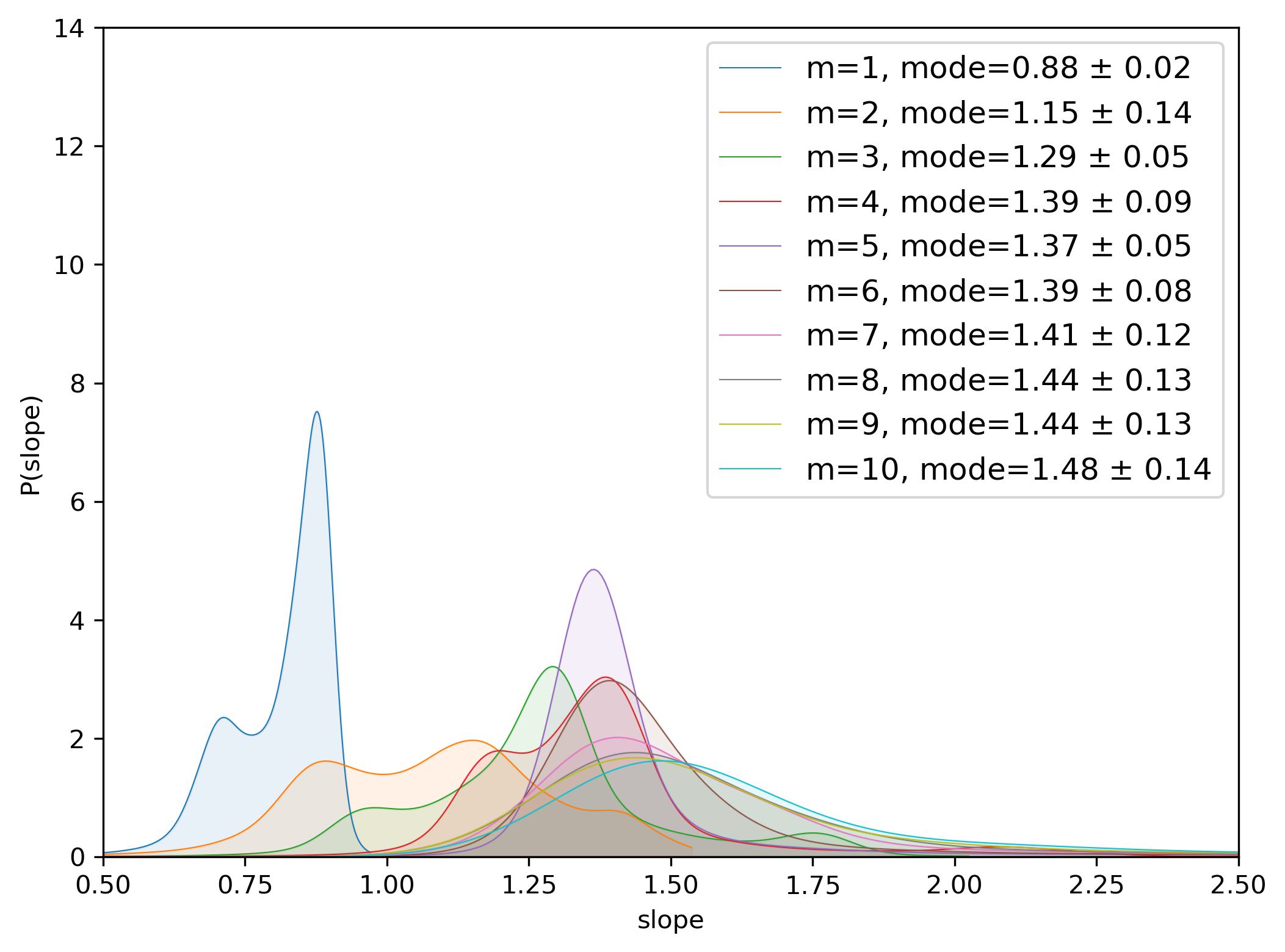}
                }
             \subfloat[Slope distributions for data set
                  \#2 ]{
                  \includegraphics[width=0.4\linewidth]{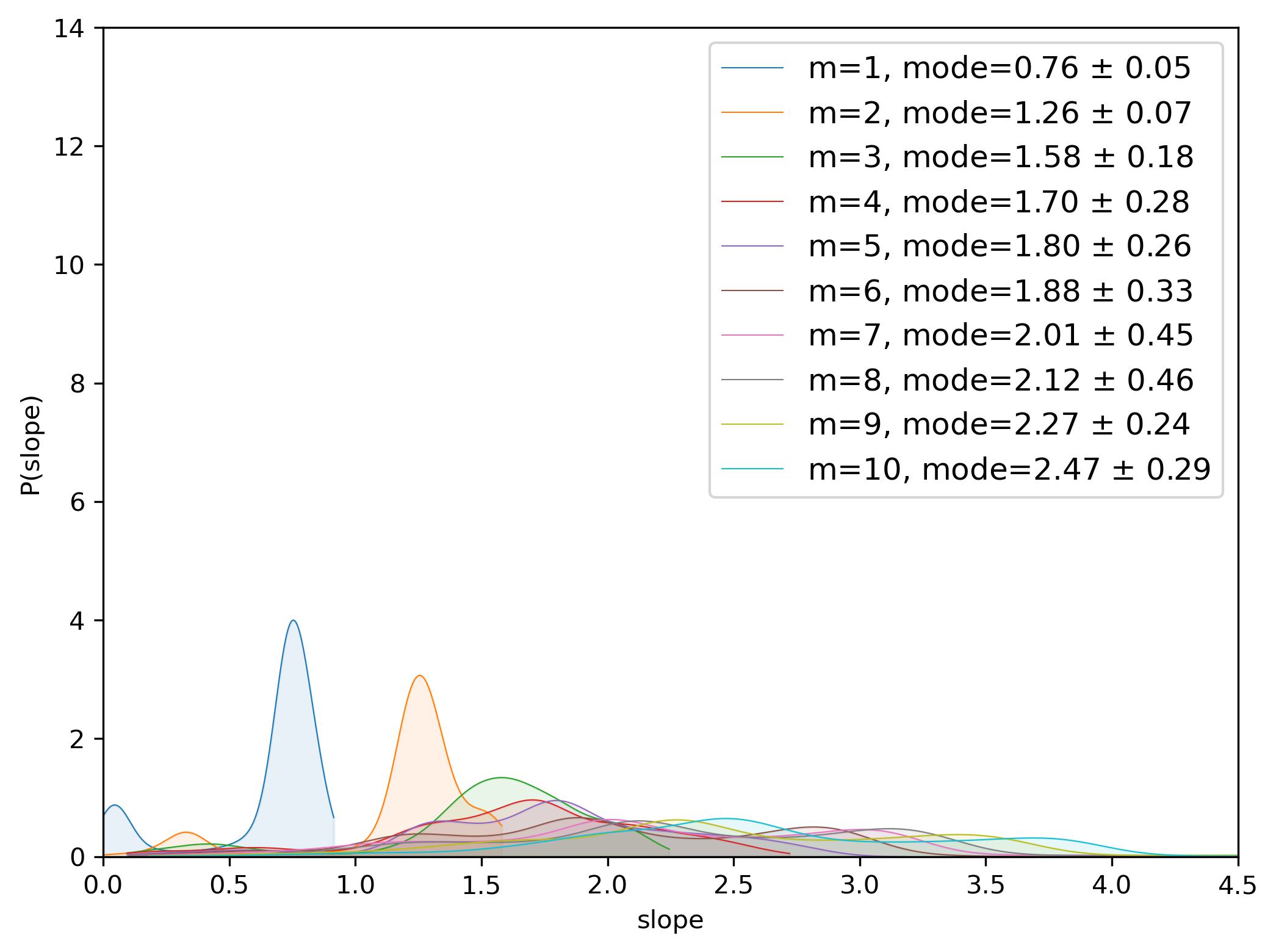}
                }\\
             \subfloat[Wasserstein distance for data
                  set \#1]{
                  \includegraphics[width=0.4\linewidth]{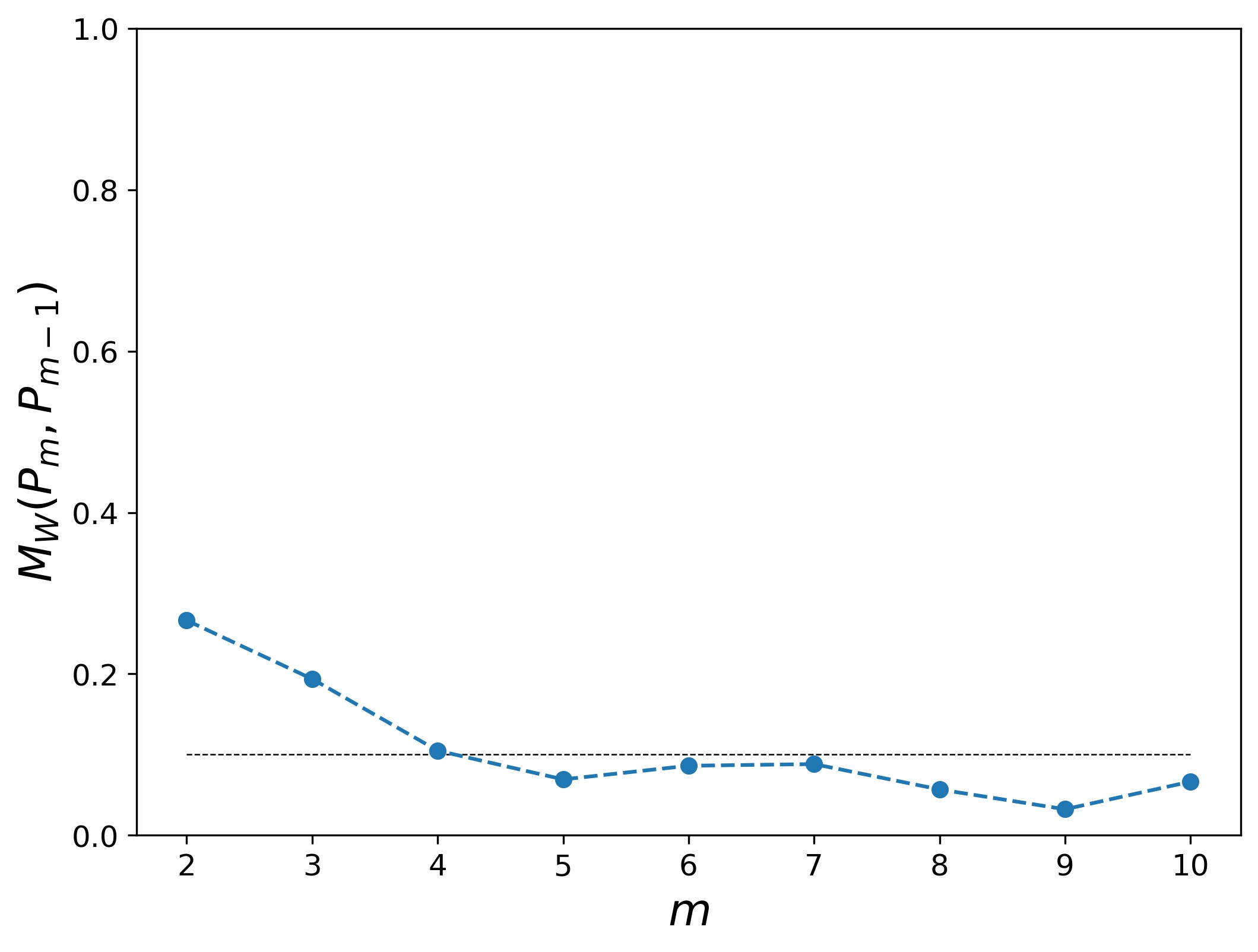}
                }
             \subfloat[Wasserstein distance for data
                  set \#2 ]{
                  \includegraphics[width=0.4\linewidth]{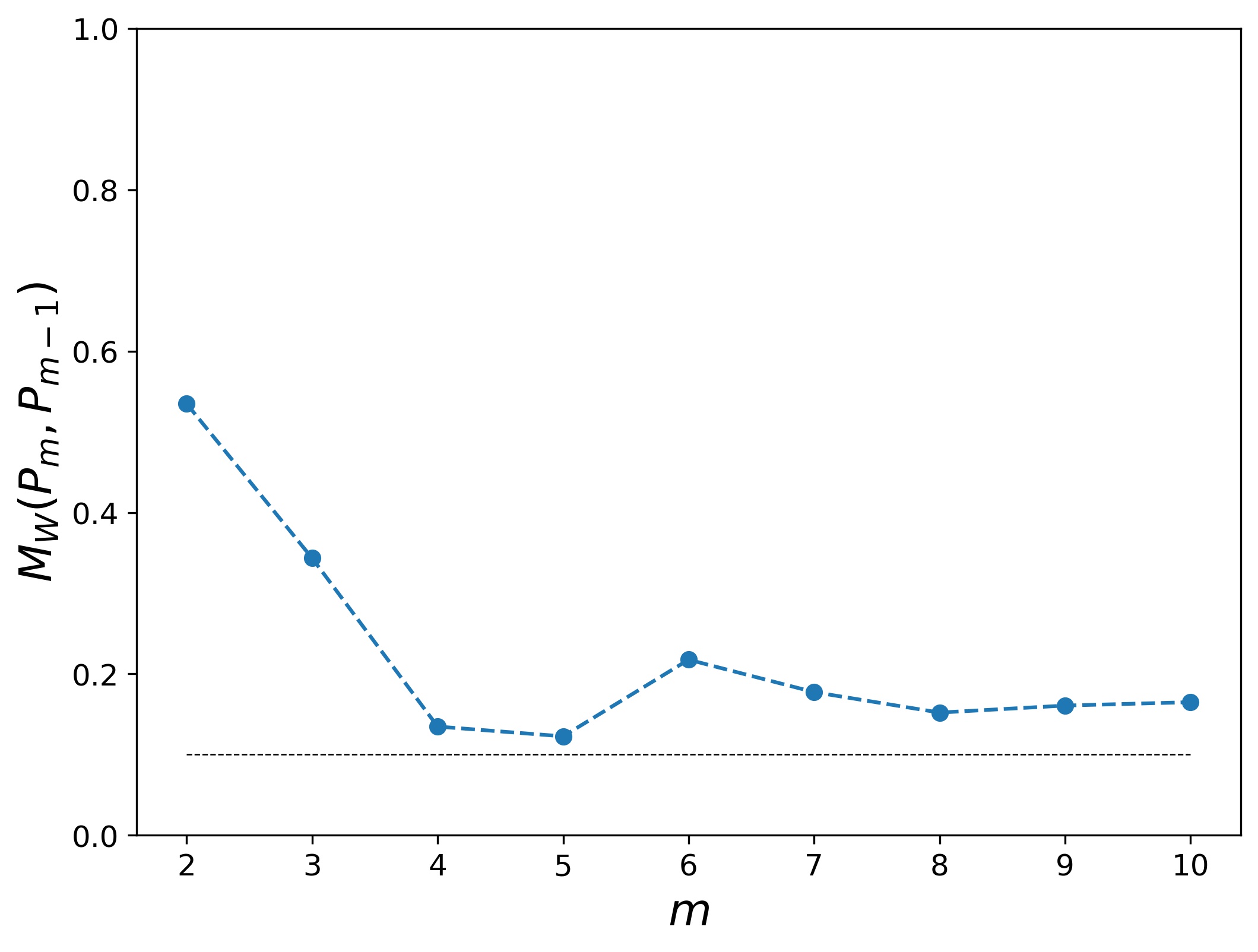}}
  \end{center}
  \caption{Extracting scaling regions for data from two laser
    experiments, segments of which are shown in the top two panels.
    Below those traces are the correlation sum plots, slope
    distributions, endpoint distributions, and convergence measures
    for the corresponding data.}
\label{fig:pic-example}
\end{figure*}

Moving beyond synthetic examples, we now consider two PIC laser
data sets from McMahon {\sl et al.}~\cite{McMahon}. These were
gathered from the same device but under different conditions and,
as noted in the paper, lead to quite different  dynamics;
see Figure~\ref{fig:pic-example}(a) and (b).
McMahon {\sl et al.} first estimate
$\tau$ using AMI
then calculate the correlation sums over a fixed range of $m\in[5,10]$.
They apply a ``minimum gradient detection'' algorithm 
to find scaling regions. This method gives $d_2 = 1.27 \pm 0.05$ and $1.01 \pm 0.06$,
respectively.  The paper does not note a ``best'' value
for $m$, as their goal is calculation of the correlation dimension,
not the embedding dimension.

The results of applying our methodology to this data are
shown in Figure~\ref{fig:pic-example}(c)-(h).  The AMI method gives
$\tau = 3$ for both cases. The correlation sum for a range
of $m$ values is shown in panels (c) and (d). Panels (e) and (f)
show the corresponding slope distributions, and (g) and (h) show the
Wasserstein distances.  For the data in the left-hand
column, the slope distributions are multimodal for $m \in [1,4]$,
reflecting the distinct linear regions in panel (c).  The PDFs in (e)
are far broader than those in Figures~\ref{fig:lorenz63-example}
and~\ref{fig:lorenz96-example},
indicating less certainty.  Nevertheless, the
Wasserstein distance in panel (g) does drop below the $0.1$ threshold
for $m = 5$, implying $d_2=1.26 \pm 0.07$.  This is in good agreement
with the quoted results of McMahon {\sl et al.}, though it should be
noted that their confidence interval is calculated differently.

The story is quite different for the second case.  The distributions
in Figure~\ref{fig:pic-example}(f) do not appear to converge with
increasing $m$; this is corroborated by the
Wasserstein metric in panel (h). Indeed, the
the curves in panel (d) are clearly
problematic from the standpoint of time-series analysis.
The $m=1$ and $m=2$ results do have scaling regions---indicated by the
strong, unimodal peaks in the blue and orange distributions in panel
(f)---but the slopes of these regions give spurious $d_2$ values
because the attractor is not reconstructed properly for such low
dimensions (as is clear from the change in slope with increasing $m$
in this range). When $m>2$, none of the $d_2$ curves have clear scaling
regions.  The slope distributions bring this out clearly: the
Wasserstein distance never falls below $0.1$,
indicating that low confidence in the
correlation dimension.  This is not in accord with the asserted
value in \cite{McMahon}, perhaps
because computing a gradient from noisy data---as is done in that
paper---is notoriously problematic.

A number of methods have been proposed to automate the estimation of
$d_2$: see, for
example, \cite{Toomey:09,CORANA2004779,casaleggio1999automatic}.
These papers essentially use following workflow: calculate
a local gradient of the correlation sum, generate a
histogram of the slopes, and then locate the peak value.  Numerical
differentiation can, of course, be problematic unless the data points
are noise free.  Our method is designed to avoid this
issue.  Since we weight the linear fits by their
length, we favor longer fits, thus de-emphasizing small-scale noise. Our
choice of the mode of the slope distribution provides a slope that 
is common to a range of endpoint choices.
Another important difference between our method and that in the cited papers 
is generality.  The primary focus of those papers is an automatic
estimate for the correlation dimension. The objective of our method is
to select a good value of the embedding dimension; the {\tt d2}
calculation is only the vehicle.  Any other dynamical invariant would be
just as good, as we show next.

\subsubsection{Other invariants}
\label{sec:other-invariants}

Correlation dimension is not the only dynamical invariant that
involves fitting a line to a scaling region.  Another important
quantity is the largest Lyapunov exponent, $\lambda_1$, as
computed by the widely used Kantz \cite{kantz94} and 
Rosenstein \cite{Rosenstein1993} algorithms. These
calculate a ``stretching factor'' $S(\Delta n)$ between nearby
trajectory points.  This also gives a scaling region to which
or method can be applied.
This, in turn, provides another opportunity for an automated
asymptotic invariant approach to choosing embedding parameter values.

\begin{figure*}
 \begin{center}
		\subfloat[Stretching factor $S(\Delta n)$]{
      			\includegraphics[height=0.22\linewidth]{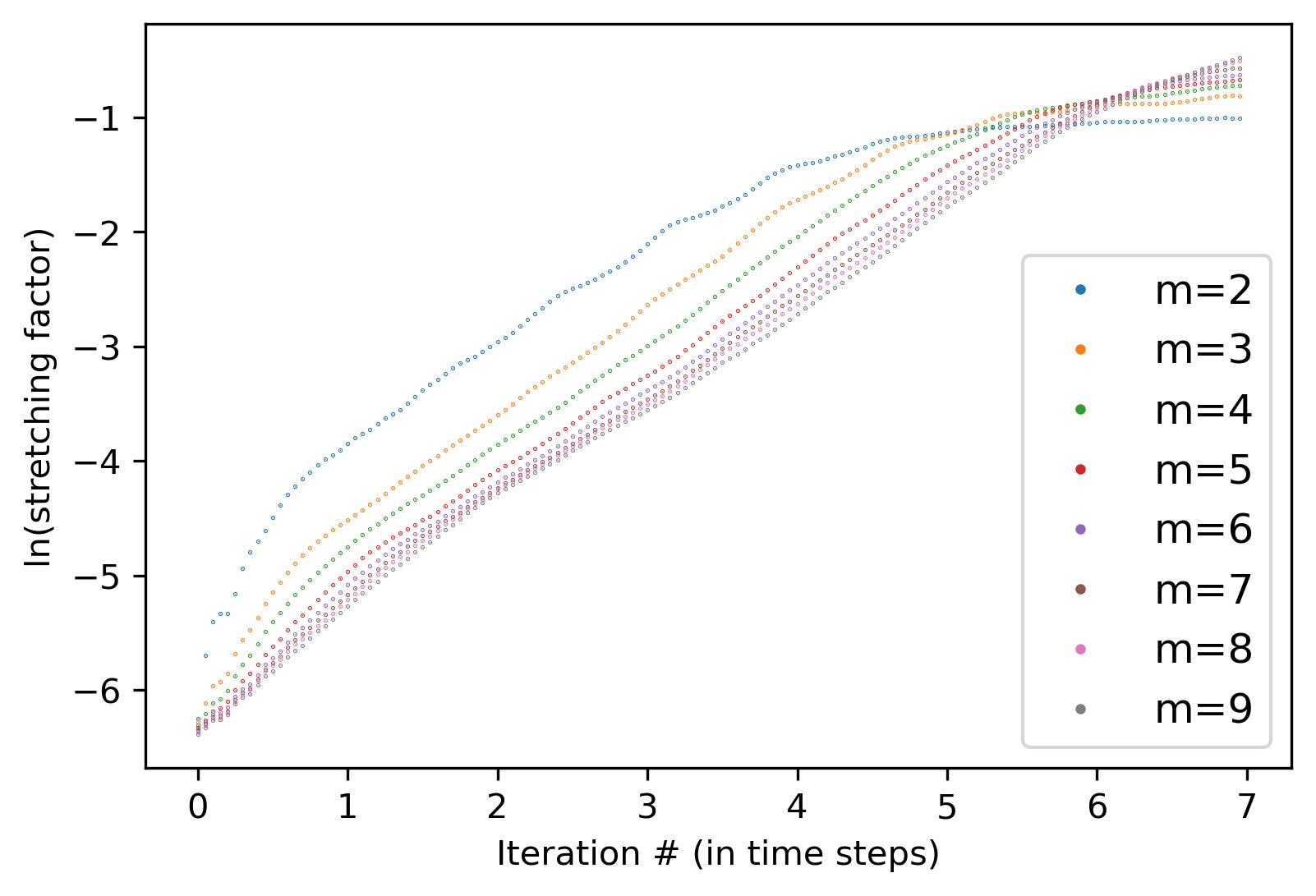} } 
      	\subfloat[Weighted slope distributions]{
                  \includegraphics[height=0.22\linewidth]{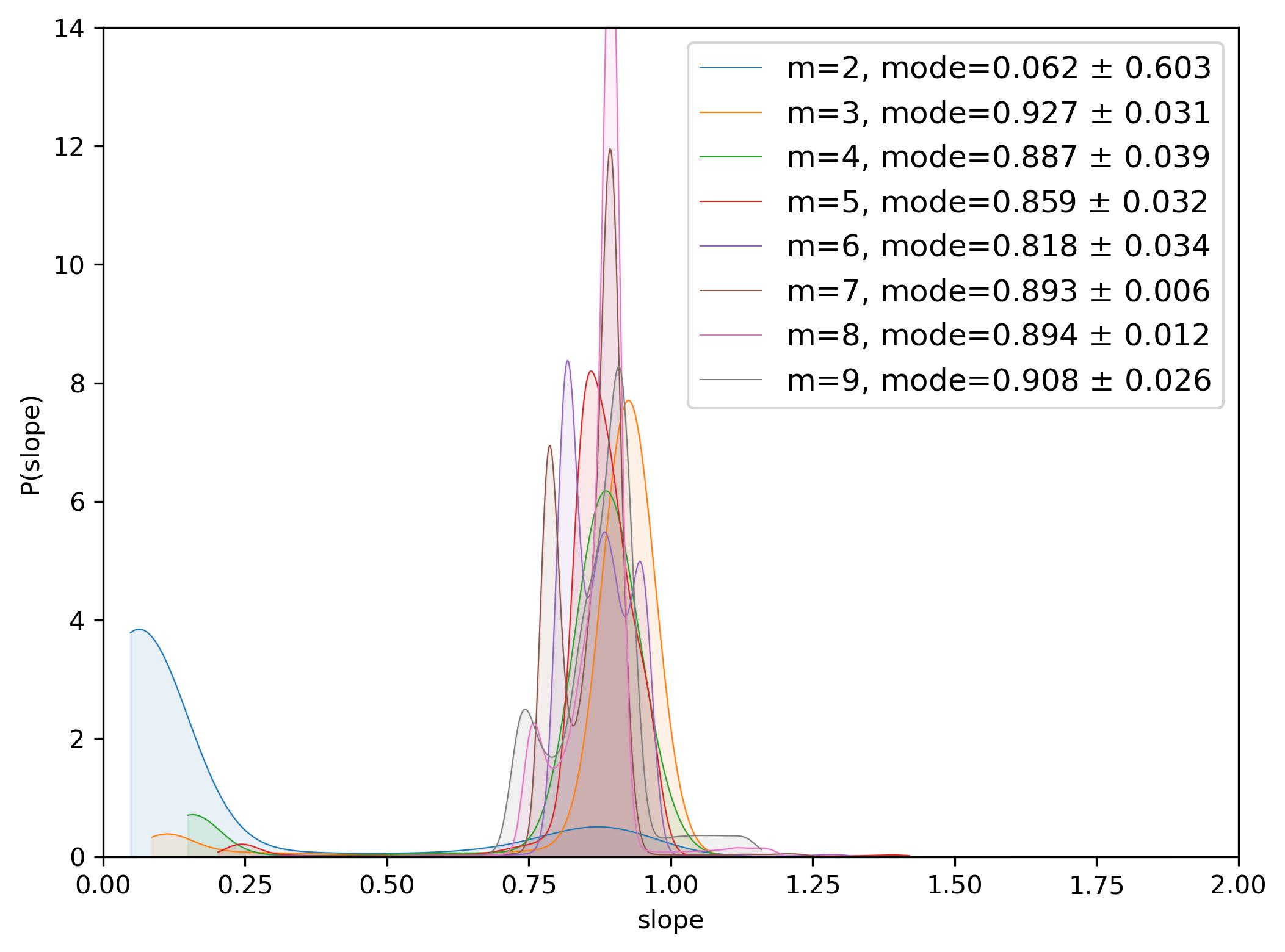}}
		\subfloat[Wasserstein distance]{
                     \includegraphics[height=0.22\linewidth]{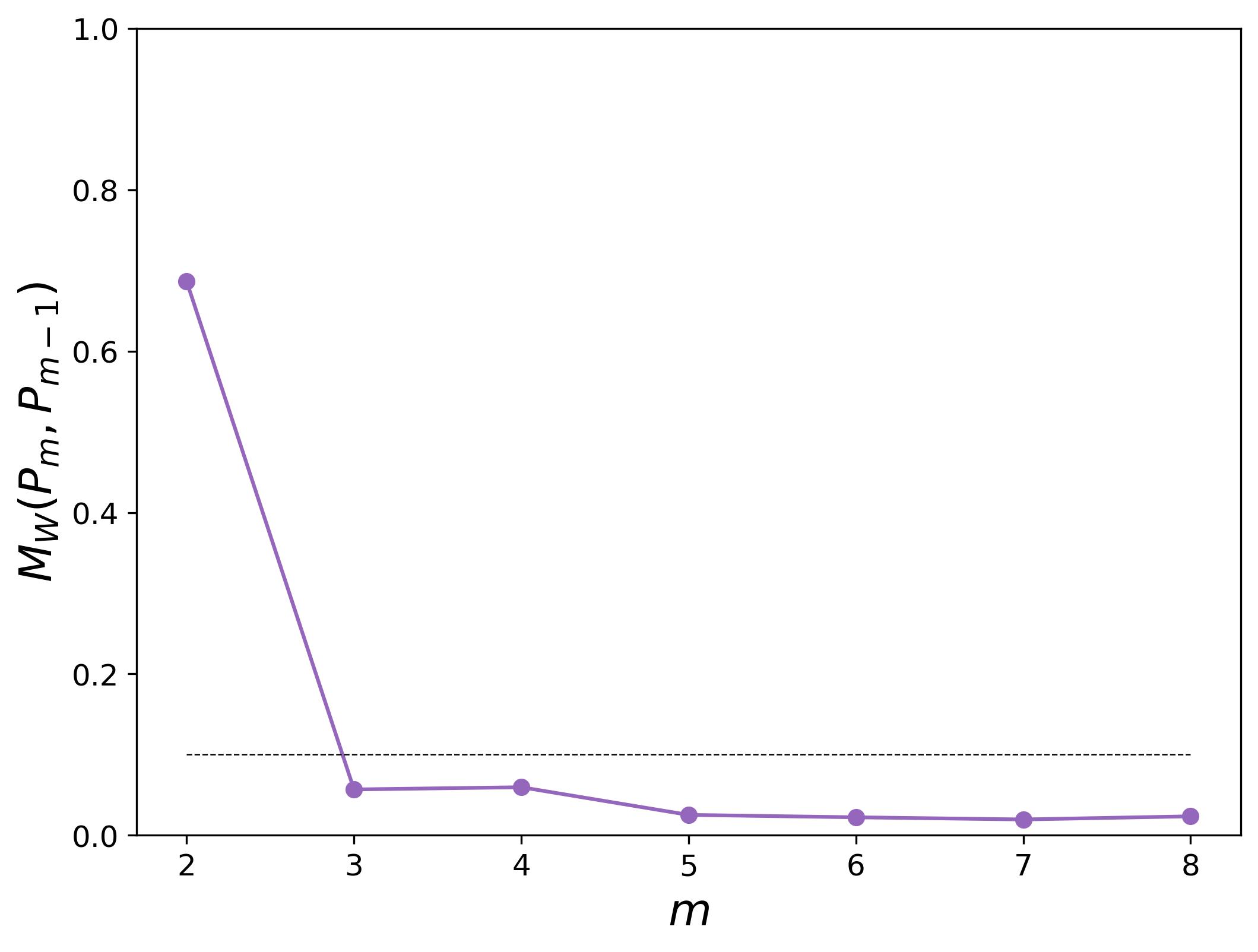}}
\end{center}
  \caption{Asymptotic invariant analysis of the Lorenz-63 system using
    the largest Lyapunov exponent: (a) spreading factor for embeddings
    for $\tau=18$ and $m=[2,9]$, computed using TISEAN's {\tt lyap\_k}
    command. (b) Weighted slope distributions generated from an
    ensemble of fits in different intervals of the plots in (a). (c)
    Convergence of slope distributions.}
\label{fig:lorenz63-lyap-example}
\end{figure*}

Figure~\ref{fig:lorenz63-lyap-example} shows the results of this
approach, as applied to the Lorenz-63 dataset from
Section~\ref{sec:datasets} using TISEAN's {\tt lyap\_k} command.
The Wasserstein plot in panel (c) suggests that $m=3$ is adequate for
this computation on this data set.  Note that this is smaller than the
value of $m=4$ that we obtained using {\tt d2} calculations on this
trajectory.  This brings out an interesting point: different values
of the embedding dimension may be sufficient for the calculation of
different invariants.  This is likely due to a combination of
both dynamical and algorithmic effects.  The {\tt lyap\_k} algorithm
analyzes how the dynamics deform the state space by tracking the
forward images of points in an initial $\epsilon$-ball that
stretch along the most unstable manifold. Our results suggest that this effect can be
tracked effectively in $m=3$, whereas the machinery of the {\tt d2}
algorithm, which counts points in $m$-dimensional $\epsilon$-balls,
requires a more fully unfolded reconstruction.  In other words, both
the nature of the invariant and the algorithm play a role.
This is not the first observation of this effect, of course,
see for example, Garland {\sl et al.}~\cite{joshua-pnp}.  

\begin{figure}
 \begin{center}
\includegraphics[width=0.5\linewidth]{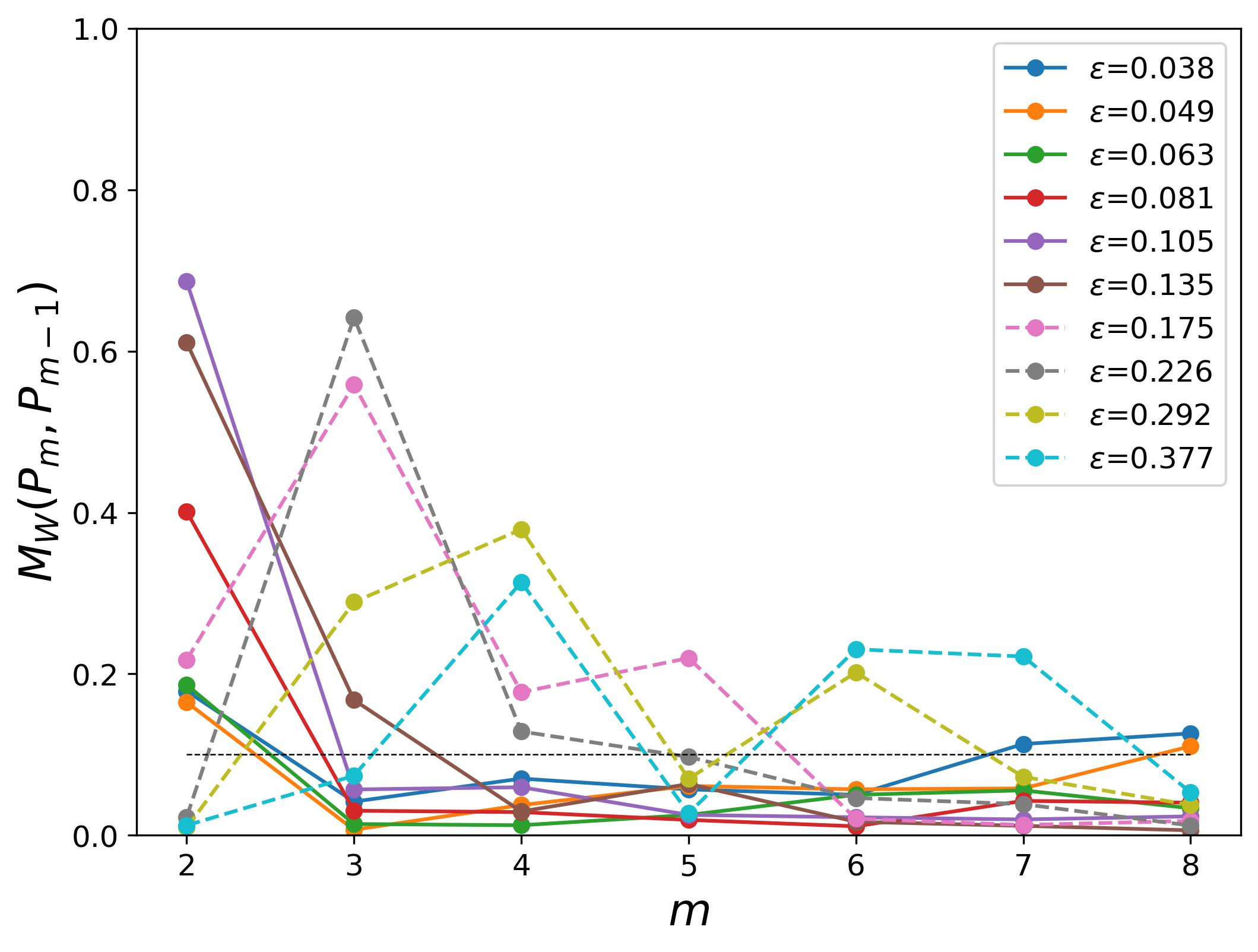}
\end{center}
 \caption{Wasserstein distance for
   the Lorenz-63 data for a range of $\epsilon$ in the Kantz algorithm for {\tt lyap\_k}.
   Figure~\ref{fig:lorenz63-lyap-example}(c) uses $\epsilon=0.105$.}
\label{fig:lorenz63-lyap-all}
\end{figure}

On a related note: the size of the initial $\epsilon$ ball in the {\tt
lyap\_k} calculation is set, by default, to five 
$1/1000$th and $1/100$th of the span of the data, and
$S(\Delta n)$  is computed for each $\epsilon$.  Data limitations can make the results quite
sensitive to this scale, however, so choosing a good
$\epsilon$ value---or knowing whether a choice is good---can
be a challenge.  Our method can provide some insight in this
situation.  Figure~\ref{fig:lorenz63-lyap-all} shows the effect of
$\epsilon$ on the Wasserstein distance for the
Lorenz-63 data.
For the five smallest values of $\epsilon$, the slopes converge by $m=3$.
For $\epsilon=0.135$, the slopes still converge, but not until $m=4$.
Beyond that, the Wasserstein distance is non-monotonic, indicating a
lack of convergence with increasing dimension. This suggests
that this range for $\epsilon$ is problematic.

\section{Discussion and conclusion}
\label{sec:conclusion}

The choice of the embedding dimension is a critical, but challenging,
step in delay reconstruction.  As discussed in the last paragraph of
Section~\ref{sec:dcorr},
a number of good heuristics have been developed to aid in this
process.
However these do not provide confidence intervals, and all involve
subjective thresholds that may or may not be optimal for any particular
data set.
In the face of this, one might adopt an iterative approach: use some
heuristic to obtain a good first guess, then computing some dynamical
invariant---e.g., the correlation dimension or Lyapunov
exponent---over a range of values for the embedding dimension and
looking for convergence.  This process, too, can be subjective, as
these computations often involve finding, and fitting a slope to, a
{\sl scaling region}. Since this is generally done by eye, it is not
immune to confirmation bias.

The contribution of this paper is a method that {\bf formalizes} and
{\bf automates} this process.  We use the ideas of Deshmukh {\sl et
  al.} \cite{varad-scaling} to generate an ensemble of slopes from
prospective scaling regions, creating a slope distribution that uses
interval width and fit quality as weights.  Broad, clean scaling
regions manifest as narrow, tall peaks in these distributions.  Upon
repeating this calculation for a range of embedding dimension, this
leads to a good choice $m$ values: when the resulting sequence of
slope distributions converges.  Convergence is signaled by the
decrease of a Wasserstein distance measure below a threshold that is
motivated by the theoretical expectation for samples from a fixed
distribution.  We demonstrated the method in Section~\ref{sec:results}
on data sets from several simulated and experimental examples. We used
two dynamical invariants calculated with the TISEAN package: the
correlation dimension and the largest Lyapunov exponent.  These are
obtained from the slope of the correlation sum versus the scale
parameter or the stretching factor versus time, respectively. The
results corroborate known values, except in one case: a laser data set
from \cite{McMahon}. In this case the correlation-sum plots, when
examined visually, clearly did not contain true scaling regions.

We emphasize that calculations of such dynamical invariants are valid
if---and only if---the plots contain ``robust'' scaling regions.
Robustness is obviously a subjective term that can lead to real
problems in the practice of nonlinear time-series analysis.  To quote
Kantz \& Schreiber: ``Some authors failed to observe that the curves
that they were fitting with straight lines were actually not quite
straight...'' \cite{kantz97}.  Fitting a line blindly to some
arbitrarily selected portion of a plot is even worse.  A strength of
our method is that it objectively measures when there is a scaling
region---and, if so, where it is, and what is its slope.

Our technique can also be useful in the {\sl invocation} of these
algorithms.  Tools like {\tt d2} or {\tt lyap\_k} in the TISEAN
package attack a difficult problem: how can one extract dynamical
invariants from incomplete samples?  These methods involve a number of
free parameters such as data length and time scale, the Theiler window
\cite{theiler-window}, etc.  The best practice for choosing such
parameters mirrors the ``asymptotic invariant'' approach: vary the
parameter, seeking convergence.  One can use our method to accomplish
this---for individual parameters or even for several at once,
using a multivariate sweep.  This could include choosing any of the free
parameters used in the practice of delay reconstruction.

\section*{Acknowledgements}
\label{sec:axk}

\noindent This material is based upon work supported by the National Science Foundation under Grants No. CMMI 1537460, CMMI 1558966, and AGS 2001670. Any opinions, findings, and conclusions or recommendations expressed in this material are those of the authors and do not necessarily reflect the views of the NSF.



\bibliographystyle{elsarticle-num} 
\bibliography{master-refs}





\end{document}